\documentclass[prl,twocolumn,superscriptaddress,nofootinbib,longbibliography,tightenlines]{revtex4-2}
\usepackage[svgnames]{xcolor}
\usepackage{amsmath,amssymb,amsthm,bbm}
\usepackage{xr-hyper}
\usepackage{hyperref}
\usepackage{graphicx}
\usepackage{float}
\usepackage{braket}
\usepackage[capitalise]{cleveref}
\hypersetup{
  colorlinks,
  citecolor=Salmon,
  linkcolor=Teal,
  urlcolor=Teal}
\usepackage[T1]{fontenc}
\usepackage[utf8]{inputenc}

\usepackage[normalem]{ulem}

\usepackage{slashed}
\usepackage{tikz}
\usepackage[compat=1.1.0]{tikz-feynman}
\usepackage{subcaption}
\usepackage{ragged2e}
\usepackage{booktabs,adjustbox}
\DeclareCaptionJustification{justified}{\justifying}
\usepackage{makecell}

\makeatletter
\newcommand*{\addFileDependency}[1]{
  \typeout{(#1)}
  \@addtofilelist{#1}
  \IfFileExists{#1}{}{\typeout{No file #1.}}
}
\makeatother

\begin{document}

\title{Quantum Monte Carlo for Gauge Fields and Matter without the Fermion Determinant}
\author{Debasish Banerjee}
\affiliation{Theory Division, Saha Institute of Nuclear Physics, 1/AF Bidhan Nagar, Kolkata 700064, India}
\affiliation{Homi Bhabha National Institute, Training School Complex, Anushaktinagar, Mumbai 400094,India}
\author{Emilie Huffman}
\affiliation{Perimeter Institute for Theoretical Physics, Waterloo, ON N2L 2Y5, Canada}
\begin{abstract}
 Ab-initio Monte Carlo simulations of strongly-interacting fermionic systems are plagued by the fermion
 sign problem, making the non-perturbative study of many interesting regimes of dense quantum matter, or
 of theories of odd numbers of fermion flavors, challenging. Moreover, typical fermion algorithms require 
 the computation (or sampling) of the fermion determinant. We focus instead on the meron cluster algorithm, 
 which can solve the fermion sign problem in a class of models without involving the determinant. 
 We develop and benchmark new meron algorithms to simulate fermions coupled to $\mathbb{Z}_2$ and $U(1)$ 
 gauge fields in the presence of appropriate four-fermi interactions. Such algorithms can be used
 to uncover potential exotic properties of matter, particularly relevant for quantum simulator 
 experiments. We demonstrate the emergence of the Gauss' Law at low temperatures for a $U(1)$ model in 
 $(1+1)-$d.
\end{abstract}

\maketitle
\paragraph{Introduction.--} Microscopic models involving fermions that strongly interact with each other,
either directly or mediated via gauge fields, are essential ingredients of many theories in condensed matter 
and particle physics \cite{Vafek2014, Borsten2017, Aitchison}. From the Hubbard model describing the physics 
of correlated fermions, to the quantum Hall effect and high-temperature superconductivity, fermions subjected 
to various interactions have been studied both perturbatively and non-perturbatively \cite{Klitzing2017,FloresLivas2020}. 
Fermions constitute the matter component of all microscopic theories of particle physics (as leptons in 
electromagnetic and weak interactions, as quarks in strong interactions) and interact with gauge fields 
(the photon, the $W^{\pm},Z$, and the gluons respectively) \cite{Peskin:1995ev}. Gauge fields are also 
becoming increasingly important to condensed matter systems, from frustrated magnetism to theories of 
deconfined quantum criticality \cite{fradkin2023field}.

  While Quantum Monte Carlo (QMC) methods are robust for non-perturbative studies of the aforementioned 
systems, they are also vulnerable to the sign problem \cite{Troyer:2004ge}. QMC methods work by performing 
importance sampling of fermion and gauge field configurations that make up the partition function. Since fermions 
anti-commute, their sign problem can be straightforwardly understood when the configurations considered are 
worldlines: whenever fermions exchange positions an odd number of times, the configuration weight acquires another 
negative sign factor, leading to huge cancellations in the summation, accompanied by an exponential increase of 
noise \cite{Banerjee:2021zed}. 

  A large family of QMC methods deal with the fermion sign problem using determinants: they introduce auxiliary 
bosonic fields and integrate out the fermions, or expand the partition function, ${\cal Z}$, as powers of the 
Hamiltonian (or parts of the Hamiltonian) and get fermion determinants for the resulting terms \cite{Blankenbecler1981, 
Assaad2008}. Because these methods result in weights that are the sums of many worldline configurations, they 
can be used more generically to simulate the largest classes of sign-problem-free Hamiltonians, with Auxiliary 
Field QMC as the most widely applicable method \cite{ALF:2020tyi, PhysRevD.82.025007, PhysRevB.71.155115,
Huffman_2014,PhysRevB.91.241117,PhysRevLett.115.250601,10.1146,PhysRevLett.116.25060,grossman2023unavoidable}. 
Determinantal methods in general scale with either the spatial lattice volume, or the spacetime lattice volume, 
which in terms of imaginary time $\beta$ and spatial lattice dimension $N$ goes either as $O(\beta^3 N^3)$ or 
$O(\beta N^3)$, depending on the method \cite{PhysRevB.34.7911,PhysRevLett.56.2521,RevModPhys.83.349,
PhysRevB.76.035116,PhysRevB.93.155117,PhysRevD.96.114502}. While this polynomial scaling is much better than the 
exponential scaling from a straightforward exact diagonalization (ED), it is much worse than the linear scaling 
achievable for spin systems, where worldline-based methods can simulate systems orders of magnitude larger than 
typical simulable fermionic systems \cite{PhysRevLett.62.361,PhysRevLett.70.875,PhysRevLett.77.5130,Kaul_2015,
sandvik2019stochastic,Sandvik:2020yup}.

 An alternative approach--well-utilized in the lattice quantum chromodynamics (LQCD) community, is the Hybrid 
Monte Carlo technique \cite{Duane:1987de,Montvay:1994cy,DeGrand:2006zz}, which computes the fermion determinant 
stochastically, and theoretically scales linearly rather than cubicly with the spatial volume. For systems of massless 
fermions and thus zero modes in ${\cal Z}$, however, the method can run into complexities and the scaling 
significantly worsens, closer to the cubic scaling of before \cite{Beyl:2017kwp}. More recently, it has been 
applied to problems in condensed matter with some promising optimizations \cite{Buividovich_2016,ulybyshev2021bridging,
lunts2022nonhertzmillis}.

  It is however possible to develop worldline-based algorithms for fermionic Hamiltonians
\cite{PhysRevB.58.4304,PhysRevB.58.R10100,PhysRevB.50.136,gattringer2022density}. 
Meron cluster methods \cite{Bietenholz_1995}, so named due to the presence of \textit{merons} (half-instantons)
in the first model for which they were developed to simulate (the 2d $O(3)$ sigma model with $\theta=\pi$), 
can be used to solve sign-problems in four-fermion Hamiltonians for certain parameter regimes, as well as for 
free fermions with a chemical potential \cite{Chandrasekharan:1999cm}. 
Because these methods sample worldlines, computing the weights scales linearly with the volume of the system, 
and negative terms in the partition function are taken care of by avoiding \textit{merons} --- this is what 
distinguishes them from bosonic simulations. The relative simplicity of these methods, with each weight 
corresponding to a worldline configuration and the lack of stabilization 
issues that can arise in determinantal methods \cite{Assaad2008}, as well as the favorable scaling of the weight 
computations, makes them an attractive choice for simulation when applicable. Correspondingly, exciting opportunities 
open up when new interesting physical models are found which can be simulated using this method
\cite{Chandrasekharan:2002vk, Assaad:2016flj, Liu:2020ygc}.

  Recently, there has been intense experimental development to study the physics of confinement and quantum spin 
liquids \cite{Martinez:2016yna, Bernien_2017, Klco_2018, Davoudi:2020yln, banerjee2021nematic, Huffman:2021gsi,
Ebadi:2020ldi} using tools of quantum simulation and computation. The microscopic models used to capture the 
physics involve fermions interacting with (Abelian) gauge fields. In this Letter, we develop meron cluster 
algorithms for a class of experimentally relevant models \cite{Chandrasekharan:1996ih,Banerjee:2012pg}, enabling 
a robust elucidation of their phase diagrams. We also introduce new classes of $\mathbb{Z}_2$ and $U(1)$ 
multi-flavored gauge-fermion theories, which might be realized in cold-atom setups and also be further studied 
using Monte Carlo techniques. Notably, the $U(1)$ family of these models seems to be one of the few families that 
falls outside the class of models known to be simulable by auxiliary-field methods, as are 
\cite{RevModPhys.83.349,Chandrasekharan:2012fk,PhysRevE.94.043311}. Moreover, the worldline nature of the method 
makes it easily employed to study the corresponding phases in these theories in higher spatial dimensions, and 
the resulting physically-relevant configurations are promising inputs for machine learning algorithms. 

\begin{figure}
\includegraphics[width=5.5cm]{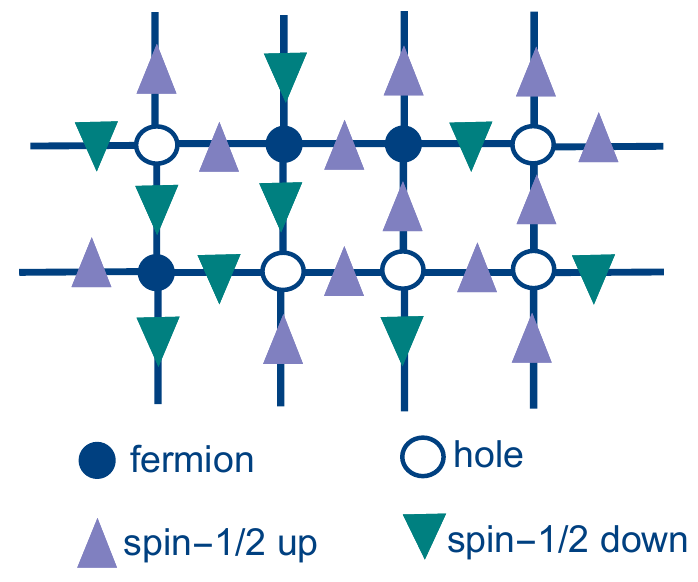}
\caption{Example fermion occupations and bond variables for the theories on a square spatial lattice with $N_f=1$.}
\label{lattice}
\end{figure}

\paragraph{Models.--}
We start with the half-filled $t$-$V$ model--a spinless fermionic Hamiltonian involving only the 
most local interactions,
\begin{equation}
H = \sum_{ \left\langle xy\right\rangle }\left[ -\frac{t}{2} (c^\dagger_x c_y + c^\dagger_y c_x) 
 + V (n_x - \frac{1}{2}) (n_y - \frac{1}{2}) \right]
\end{equation} 
 Here $\left\langle xy\right\rangle$ are the nearest neighbor site pairs, $c^\dagger, c$ are the creation and annihilation operators 
 respectively, and the repulsive interaction $V$ is given in terms of the number operator $n = c^\dagger c$. 
 It is simulable by meron clusters for $V\geq 2t$ 
 \cite{Chandrasekharan:1999cm, Liu:2020ygc}. In this Letter, we extend the meron cluster 
 method to physically interesting Hamiltonians involving gauge fields, which are lower-dimensional versions 
 of quantum electrodynamics 
 (QED) \cite{Zache:2021ggw,Hashizume:2021qbb}.
The $\mathbb{Z}_2$- and $U(1)$-gauge symmetric families are given by:
\begin{equation}
    H^{(g)}_{N_f} = -\sum_{\left\langle xy\right\rangle} \prod_{f=1}^{N_f} \left(H^{(g)}_{\left\langle xy\right \rangle, f} + H^{(g),\mathrm{des}}_{\left\langle xy\right\rangle, f}\right)
    \label{eq:modelfam}
\end{equation}
The label $g\in\left\{U(1),\mathbb{Z}_2\right\}$ is the gauge symmetry, with
\begin{equation}
\begin{aligned}
    H^{\mathbb{Z}_2}_{\left\langle xy\right\rangle, f} &=  
    t\left(c^\dagger_{x,f}s^1_{xy,f}c_{y,f} +c^\dagger_{y,f}s^1_{xy,f}c_{x,f}\right)\\ 
    H^{U(1)}_{\left\langle xy\right\rangle, f} &= 
    t\left(c^\dagger_{x,f}s^+_{xy,f}c_{y,f} +c^\dagger_{y,f}s^-_{xy,f}c_{x,f}\right).
    \end{aligned}
    \label{eq:local}
\end{equation}
The hopping of spinless fermions between the nearest neighbors $\left\langle xy\right\rangle$ are now governed by the presence
of gauge fields, represented by spin-1/2 operators, $s^k_{xy}$, on the bond. \cref{lattice} illustrates the model degrees of freedom.
Then $H^{(g),\mathrm{des}}_{\left\langle xy\right\rangle, f}$ is a designer term \cite{Kaul_2013} that 
makes the models simulable by the meron algorithm,
\begin{equation}
    \begin{aligned}
   H^{\mathbb{Z}_2,\mathrm{des}}_{\left\langle xy\right\rangle, f} &= 
   -2t\left(n_{x,f}-\frac{1}{2}\right)\left(n_{y,f}-\frac{1}{2}\right)+ \frac{t}{2}\\
   H^{U(1),\mathrm{des}}_{\left\langle xy\right\rangle, f} &= 
   -t\left(n_{x,f}-\frac{1}{2}\right)\left(n_{y,f}-\frac{1}{2}\right)\\   
    & \qquad\qquad\quad-t s_{xy,f}^3\left(n_{y,f}-n_{x,f}\right) + \frac{t}{4}
    \end{aligned}
    \label{eq:design}
\end{equation}

\begin{figure}[t]
    \centering
    \includegraphics[width=1.1in]{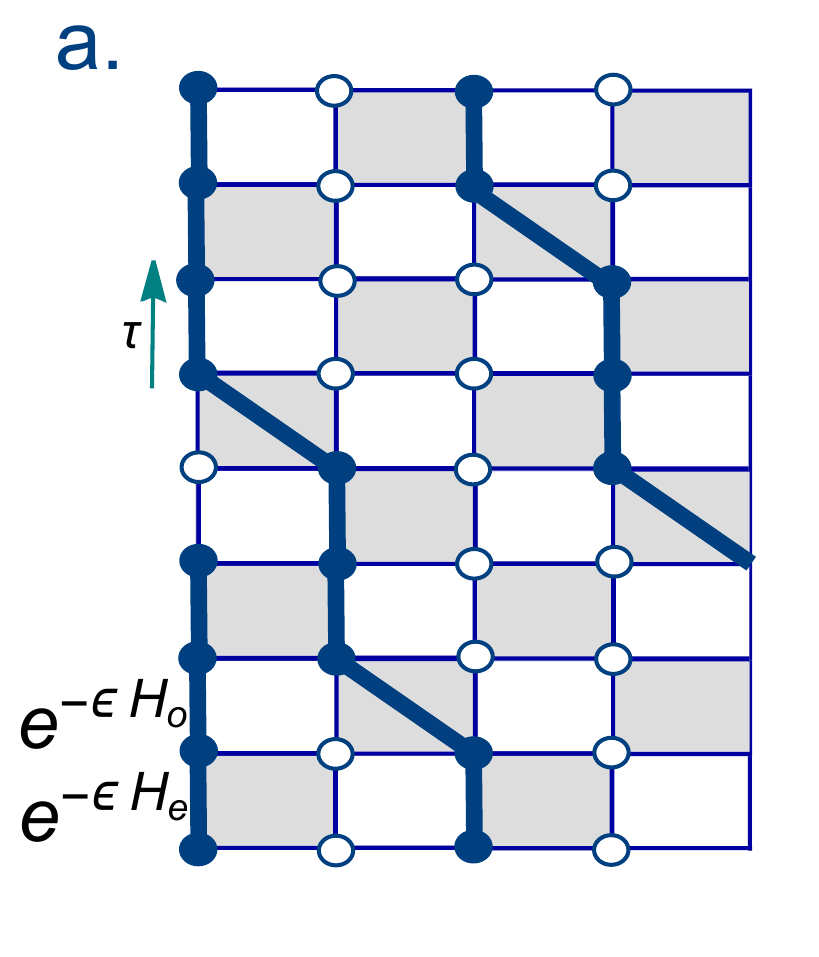}
    \includegraphics[width=1.1in]{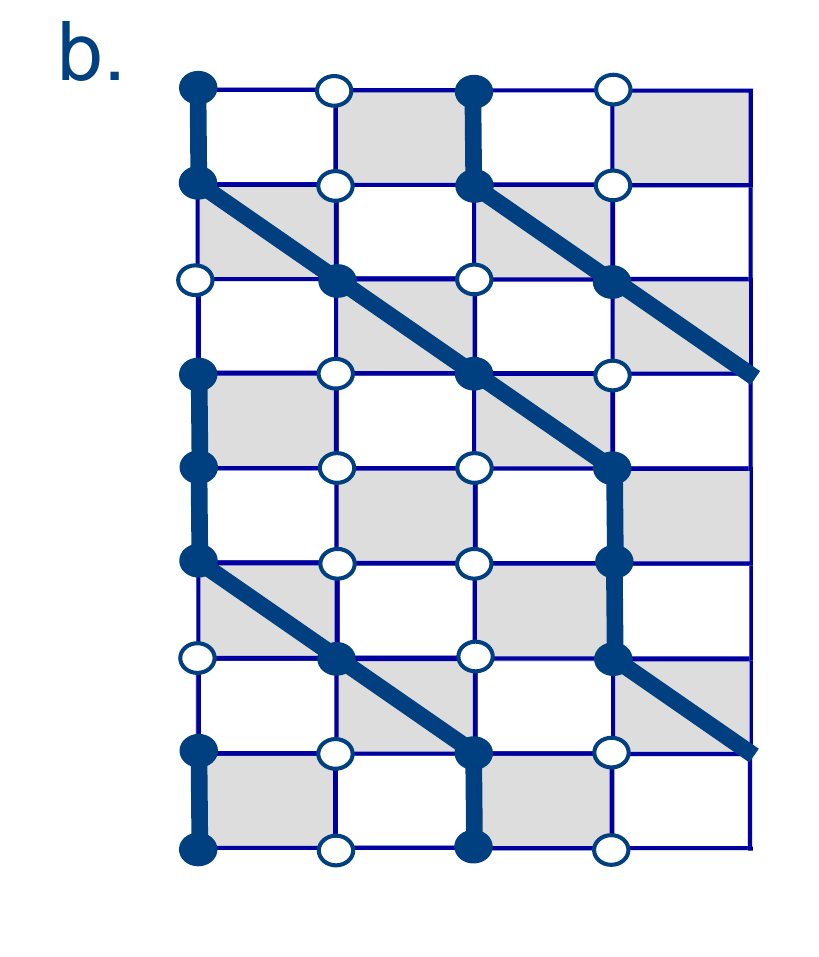}
    \includegraphics[width=1.1in]{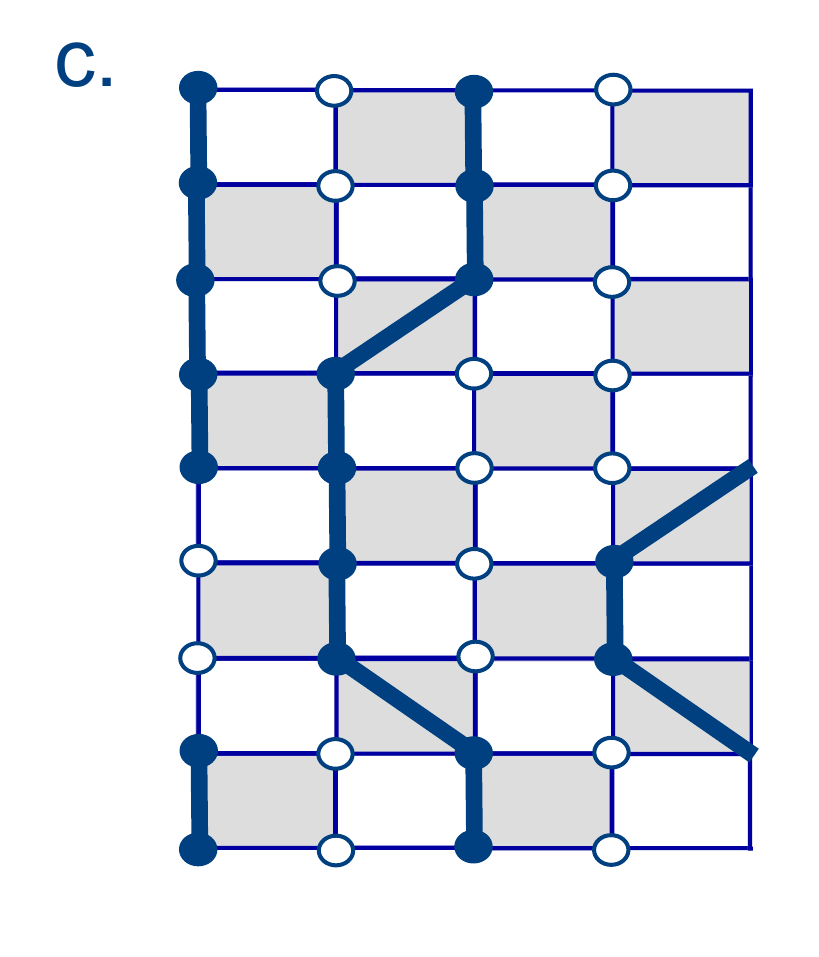}
    \includegraphics[width=1.1in]{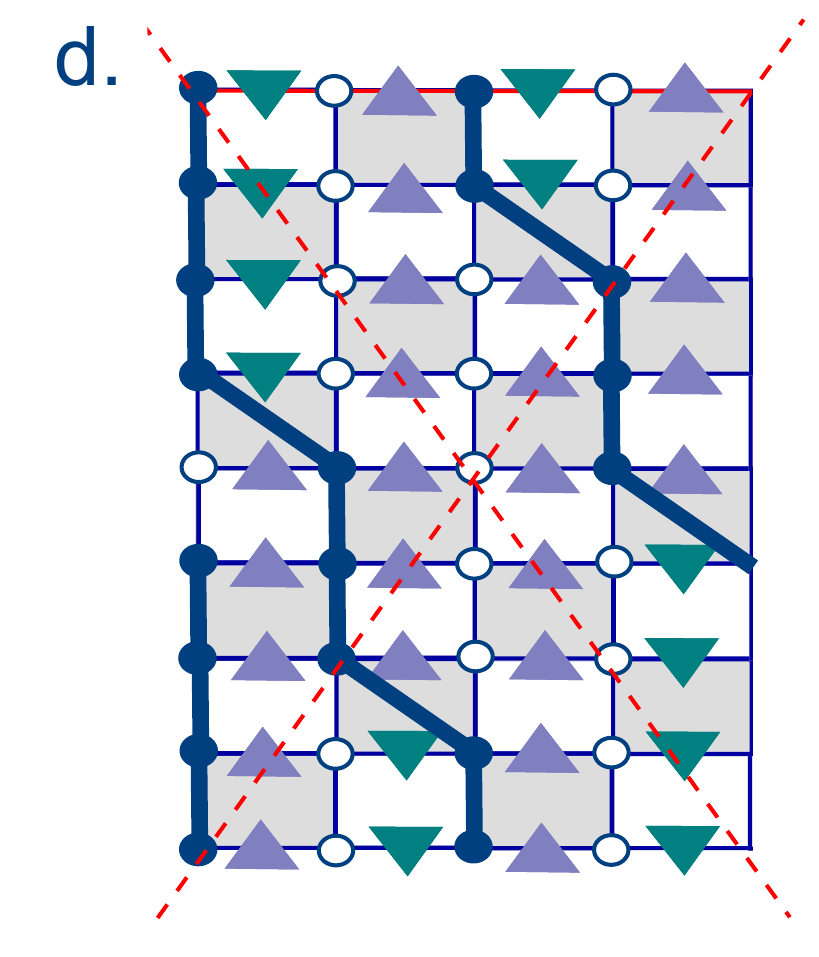}
    \includegraphics[width=1.1in]{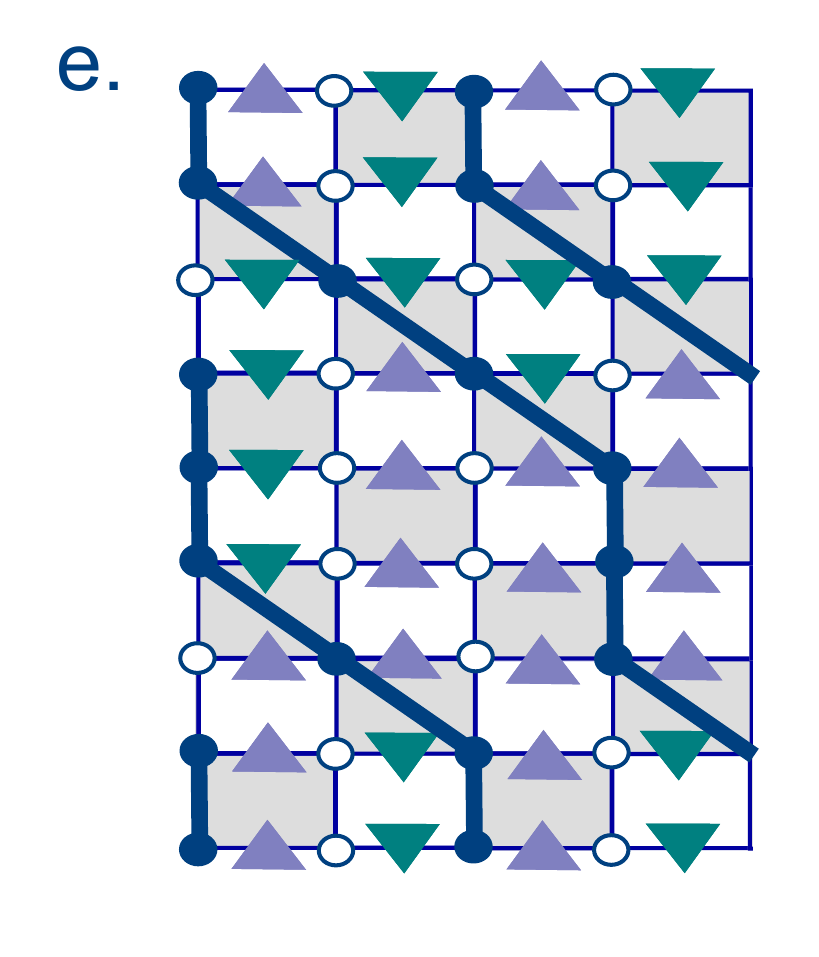}
    \includegraphics[width=1.1in]{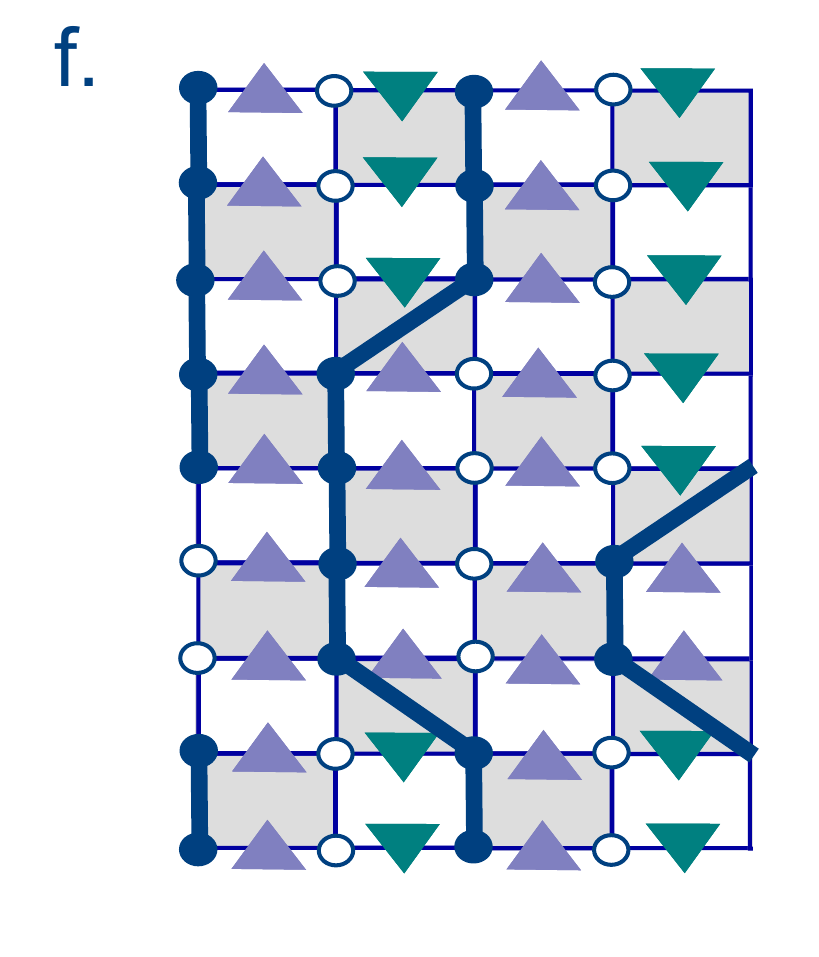}
    \includegraphics[width=1.1in]{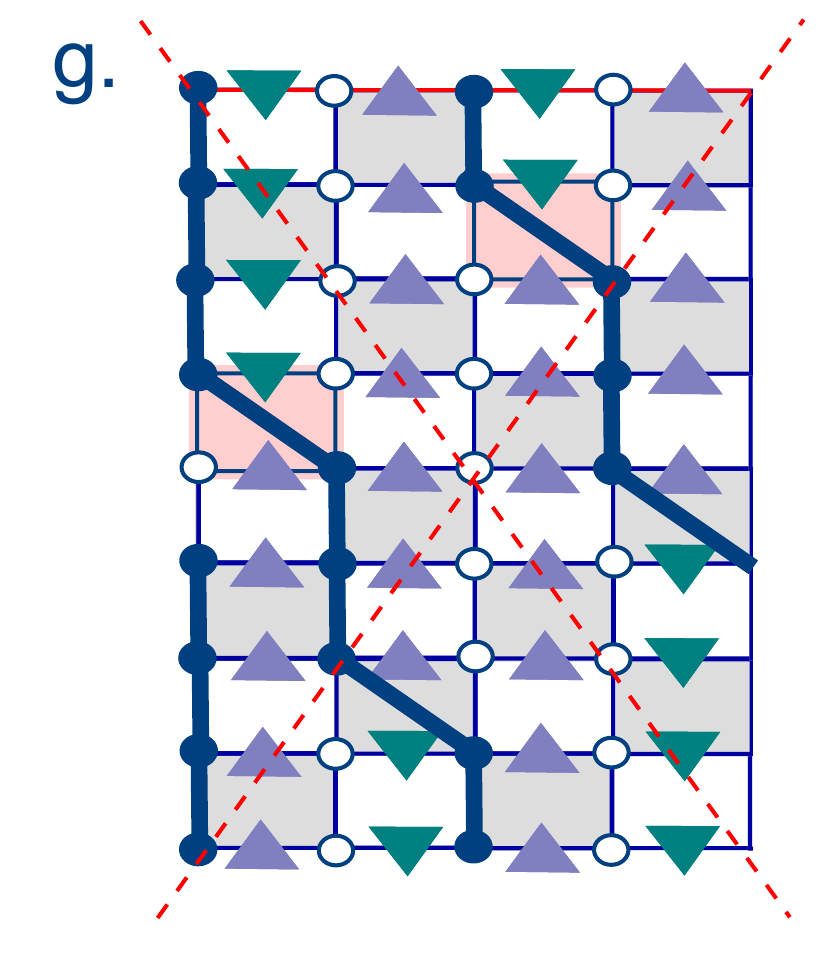}
    \includegraphics[width=1.1in]{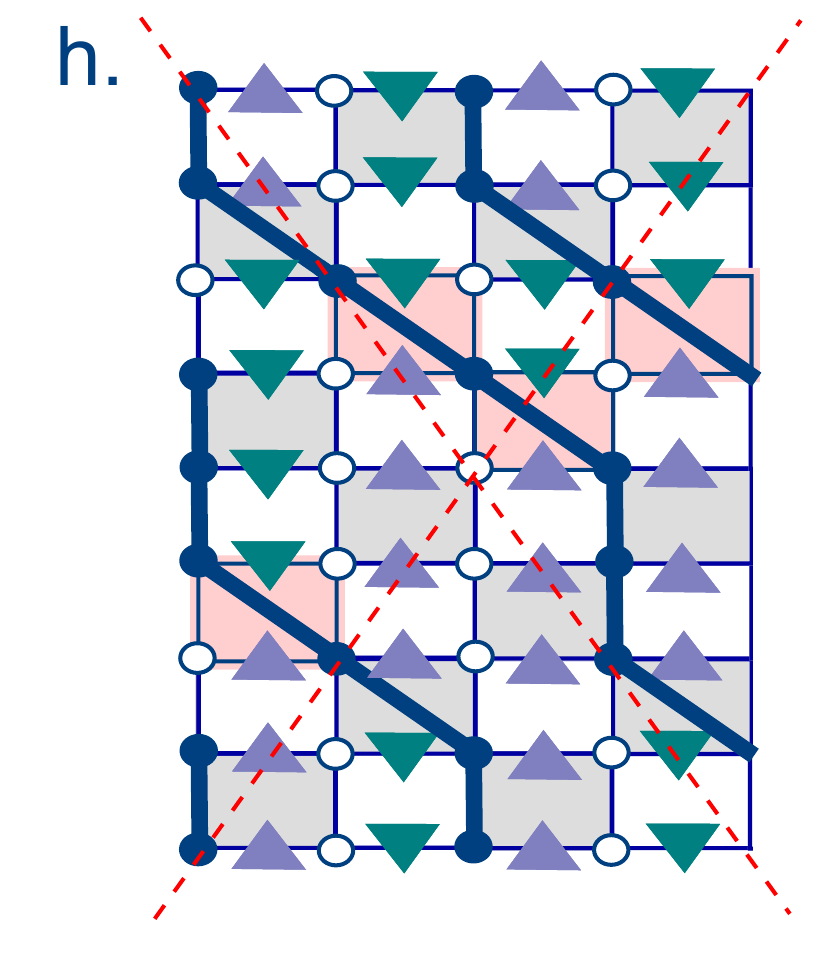}
    \includegraphics[width=1.1in]{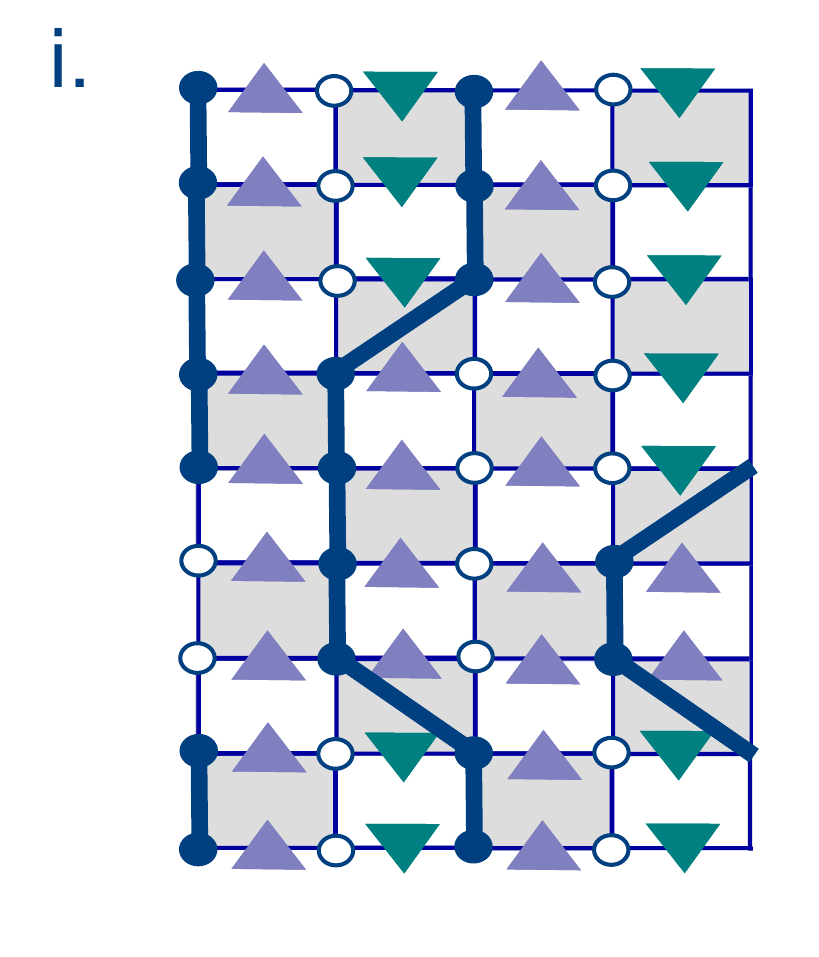}
    \caption{Worldlines for the $t$-$V$ model are in (a)-(c), the $\mathbb{Z}_2$ theory in (d)-(f), and the $U(1)$ 
    theory in (g)-(i). Image (a) shows the imaginary time direction and the $(1+1)$-d trotterization, which is the 
    same for all images. Filled circles are sites occupied by fermions, and empty circles are holes. Figures in the 
    second two rows also have link variables because they correspond to gauge theories: the upward triangles correspond 
    to spin $+1/2$ and the downward triangles correspond to spin $-1/2$. While the fermionic worldlines are the same 
    in each column, some configurations that are allowed for the $t$-$V$ model have zero weight for the $\mathbb{Z}_2$ 
    and $U(1)$ theories. These are crossed out, and zero-weight plaquettes are shaded red.}
    \label{fig:worldlines}
\end{figure}

These terms serve a similar role as the $V=2t$ term in the meron cluster 
algorithm applied to the $t$-$V$ model \cite{Chandrasekharan:1999cm}, and similarly an additional particle-hole 
symmetric $V\geq 0$ term can be added to the models here in a sign-problem-free way. In the 
$\mathbb{Z}_2$ gauge theory, the gauge field $s^1=\sigma^1/2$ couples to fermions, and 
the local $\mathbb{Z}_2$ symmetry is manifest via the commutation $[Q_x,H^{\mathbb{Z}_2}]=0$,
where $Q_x = (-1)^{\sum_f n_{x,f}}\prod_{f,\hat{\alpha}} s^3_{x,x+\hat{\alpha},f}  s^3_{x-\hat{\alpha},x,f}$, 
$\hat{\alpha}$ are the unit vectors in a $d$-dimensional square lattice. For the $U(1)$ theory, 
the unitary operator $V_{U(1)}$, which commutes with $H^{U(1)}$ is given by 
$V_{U(1)} = \prod_x e^{i\theta_x G_x}$, with $G_x = \sum_{f} \left[n_{x,f} - 
    \sum_{\hat{\alpha}}\left(s^3_{x,x+\hat{\alpha},f}- s^3_{x-\hat{\alpha},x,f}\right)+((-1)^x-1)/2\right]$.
 In the terminology of gauge fields, our microscopic models are \emph{quantum link models} \cite{Wiese:2021djl}, 
which realize the continuous gauge invariance using finite-dimensional quantum degrees of freedom. The identification 
with usual gauge field operators is $U_{xy,f} = s^+_{xy,f},~ U^\dagger_{xy,f} = s^-_{xy,f}, E = s^3_{xy,f}$.
 We note that a straightforward application of the meron idea necessitates the introduction of an equivalent 
\emph{flavor} index for gauge links as fermion flavors. Naively, the total Gauss law can be expressed through a 
product ($\mathbb{Z}_2$) or sum ($U(1)$) of the Gauss law of individual flavors degrees of freedom, and the resulting 
theories have $\mathbb{Z}_2^{\otimes N_f}$ and $U(1)^{\otimes N_f}$ gauge symmetry. However, flavored gauge-interactions 
can also be turned on in the $U(1)$ model (as explained in the Supplementary Material),
\begin{equation}
\begin{aligned}
 H^{U\left(1\right)}_{N_f=2} \rightarrow & 
 H^{U\left(1\right)}_{N_f=2} + J\sum_{\left\langle xy\right\rangle} s^3_{xy,1} s^3_{xy,2},
    \end{aligned}
    \label{extra}
\end{equation}
or through a Hubbard-U interaction for both $\mathbb{Z}_2$ and $U(1)$-symmetric models \cite{Liu:2020ygc}. 
These additions directly cause ordering for either the gauge field or the fermions, and the coupling 
between them leads to the interesting question of how the other particles are affected by this ordering. In 
similar contexts, interesting simultaneous phase transitions of both the fermions and gauge fields have been 
found \cite{Assaad:2016flj,Gazit_2017,Gazit_2018,PhysRevX.9.021022,Borla_2020,Frank_2020,Hashizume:2021qbb}
or conjectured \cite{homeier2022quantum, PhysRevD.19.3682}.

\paragraph{Algorithm.--}
 The algorithm is best understood through the worldline configurations for the 
 models defined in \cref{eq:modelfam,eq:local,eq:design} in the occupation number basis for the 
 fermions and the electric flux (spin-$z$) basis for the gauge links. The partition function in $(1+1)$-d 
 is
\begin{equation}
\begin{aligned}
    &{\cal Z} ={\mathrm{Tr}} \left(e^{-\beta H}\right)\\
    &\; = \sum_{\left\{s,n\right\}} \left\langle s_1,n_1\right| e^{-\epsilon H_e}\left|s_{2N_t},n_{2N_t}\right\rangle\left\langle s_{2N_t},n_{2N_t}\right|\\ &\qquad\times e^{-\epsilon H_o}...e^{-\epsilon H_e}\left|s_2,n_2\right\rangle\left\langle s_2,n_2\right|
    e^{-\epsilon H_o}\left|s_1,n_1\right\rangle ,
    \end{aligned}
    \label{eq:worldlines2}
\end{equation}
where $H = H_e + H_o$, and $H_e$ ($H_o$) consists of Hamiltonian terms that correspond to even (odd) links. 
This Trotterized approximation, is a sum of terms over discrete time-slices $1, \cdots, 2N_t$, 
each with locally defined electric flux and fermion occupation numbers. All terms within $H_e$ and $H_o$ 
commute with each other (there are straightforward generalizations for higher dimensions) 
\cite{PhysRevB.33.6271}. Each of the terms in \cref{eq:worldlines2} is a worldline configuration, 
and the rules for allowable worldline configurations apply consistently to all models within each of 
the symmetry families. \cref{fig:worldlines}(a)-(c) give examples of such configurations 
for the $t$-$V$ model as simulated by meron clusters. 

In the $\mathbb{Z}_2$ case, for each time-slice a fermion has the option of hopping to an unoccupied nearest 
neighbor site of the same flavor. 
The hop flips the flux on the bond between the sites of the same flavour index--this is 
the result of the $s_{xy}^1$ operator. \cref{fig:worldlines}(d)-(f) gives example configurations for the 
$N_f=1$ version of this model. Due to the trace condition, odd winding numbers 
are ruled out because these would cause mismatch between the spins in the initial 
and the final state.

 The possible worldline configurations for the $U(1)$ case are even more restrictive than  
the $\mathbb{Z}_2$ case. The $s^+_{xy}$ and $s^-_{xy}$ operators allow the hopping for a given flavor of 
fermion only in one direction or the other for each bond, depending on the orientation of the same flavored 
flux on the bond. \cref{fig:worldlines}(g)-(i) illustrates an example configuration and restrictions for the 
single flavor version of the $U(1)$ model. In $(1+1)$-d, it is clear that all allowed configurations must 
have zero winding number.

The worldline configurations are a tool to obtain meron cluster configurations by introducing appropriate breakups, 
which decompose the terms in \cref{eq:worldlines2} into further constituents. In considering the allowed worldline 
configurations given in \cref{fig:worldlines} for the $U(1)$ theory, for example, each of the active plaquettes in each 
time-slice (shaded in gray) must be one of the plaquettes given in Table \ref{tab:first}. The plaquettes in each row 
share the same weight, computed using $\left\langle s_b, n_b\right| e^{-\epsilon H_b}\left|s_b', n_b'\right\rangle$, 
from \cref{eq:worldlines2}, where $b$ is a nearest neighbor bond, $b=\{x,y\}$. The corresponding breakup cell for each 
row gives allowable breakups: if all fermion occupations/spins are flipped along any one of the lines, the resulting 
plaquette also exists in this table. From the table, we see two such breakups are defined, $A$ and $D$. 
Although these breakups resemble those from the original meron algorithm, in our case the breakups
involve the link variables as well--either as additional lines for the $A$ breakups, or as binding lines extending 
outward from the horizontal $D$ breakup lines. This is a key difference for the gauge extension of the algorithm. 
By computing the matrix elements that correspond to the plaquettes in each grouping, we find that for the $U(1)$ 
theory, the corresponding breakup weights $w_A$ and $w_D$ must obey:
\begin{equation}
    \begin{aligned}
    w_A &= 1 \\
    w_D &= \exp \left(\epsilon t\right) \sinh \epsilon t \\
    w_A + w_D &= \exp\left(\epsilon t\right) \cosh \epsilon t ,
    \end{aligned}
\end{equation}
\begin{table}
    \centering
    \begin{tabular}{ m{5.45cm} | m{2.7cm}}
    \hline
        Plaquettes & Breakups  \\
         \hline
         \includegraphics[width=1.3cm]{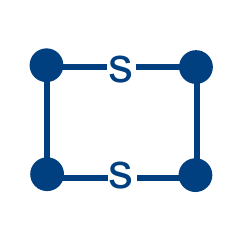} 
         \includegraphics[width=1.3cm]{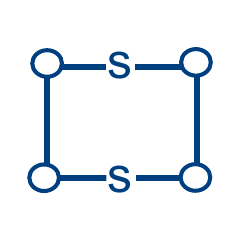} 
         \includegraphics[width=1.3cm]{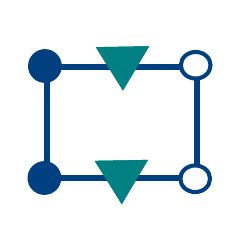} 
         \includegraphics[width=1.3cm]{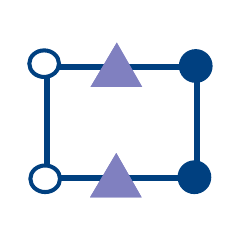}   &
         
         \includegraphics[width=1.3cm]{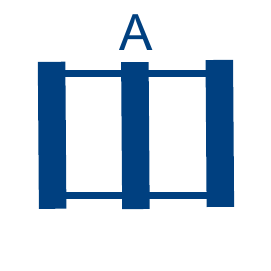}
         \\
         \hline
         \includegraphics[width=1.3cm]{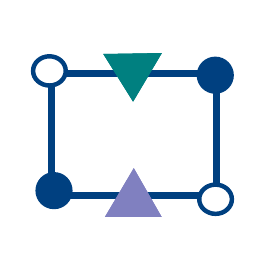}\includegraphics[width=1.3cm]{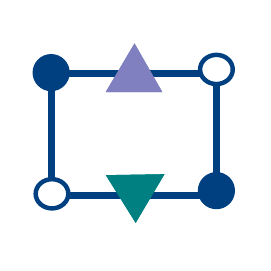} &
         \includegraphics[width=1.3cm]{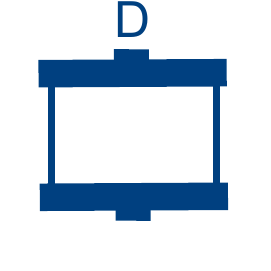} \\
         \hline
         \includegraphics[width=1.3cm]{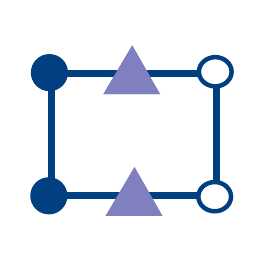}\includegraphics[width=1.3cm]{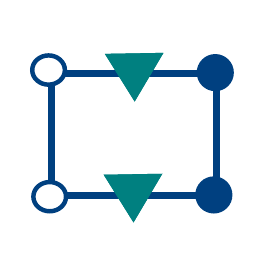} &
         \includegraphics[width=1.3cm]{Figures/pa.pdf}  \includegraphics[width=1.3cm]{Figures/dbreakup2.pdf}    
         \\
         \hline
    \end{tabular}
    \caption{Plaquettes and breakups for the $U(1)$-symmetric Hamiltonian. The middle 
    cluster lines in the $A$-breakups and binding lines in the $D$-breakups distinguishes them from 
    the original meron cluster breakups.}
    \label{tab:first}
\end{table}
to satisfy detailed balance. Moreover, the choice of the breakups is such that the total sign of a configuration 
factorizes into a product of the signs of each cluster: 
Sign $\left[C\right]$ = $\prod_i^{N_c}$ Sign$\left[ C_i\right]$, where the configuration $C$ has been decomposed
into $N_c$ clusters. We can thus simulate this system by exploring a configuration space where each configuration is defined according to the 
combination of worldlines and breakups. By assigning breakups to all active plaquettes, clusters are formed, and then 
updates involve flipping all fluxes and fermions within a cluster, which generates a new worldline configuration. 
The algorithm begins with putting the system in a \textit{reference configuration}, 
defined by the fermionic worldlines only, where the weight is known to be positive, and it is 
always possible to reach this configuration by appropriately flipping a subset of clusters
in a given configuration. For both $U(1)$ and $\mathbb{Z}_2$ theories, the reference configuration has a 
staggered fermionic occupation (charge density wave, or CDW), where fermions 
and fluxes are stationary throughout imaginary time. Fluxes can be in any spatial configuration (because they do 
not contribute a sign), and the breakups are all A. Fluxes and breakups may be initially attached to the plaquettes 
in any way allowed by Table \ref{tab:first}. A QMC sweep is then:
\begin{enumerate}
    \item Go through the list of the active plaquettes and update each breakup, one at a time.
    \begin{enumerate}
        \item If the breakup can be changed for a plaquette, change it with probability dependent on the breakup weights.
        \item If the breakup is changed, consider the new configuration that would result from this change. If it contains 
        a cluster where flipping the fermion occupation causes the fermions to permute in a way that produces a negative 
        sign, then it is a \textit{meron}. In that case, restore the breakup back to its initial state. 
        Rules for identifying merons generalize \cite{Chandrasekharan:1999cm, Chandrasekharan:2000fr} and are in the 
        Supplementary Material.
    \end{enumerate}
    \item Identify the new clusters formed by the breakups in the new configuration. For each cluster, flip all 
    fermions and fluxes with probability $1/2$.
\end{enumerate}
This describes sampling of the zero-meron sector only, but sectors with other numbers of merons may become relevant 
depending on the observable \cite{Chandrasekharan:1999cm}. We note that the cluster rules implement the Hamiltonian 
dynamics, but the constraints due to Gauss' law are not included. Like any cluster algorithm, once the 
detailed balance conditions have been satisfied, the meron algorithm is expected to be efficient in any space-time 
dimension \cite{Chandrasekharan:2002vk,Chandrasekharan:2000fr}. 
We provide a demonstration of the efficiency in $(2+1)-$d in the Supplementary Material, with 
an extensive investigation left for future work. 

\begin{figure}
    \centering
    \includegraphics[scale=0.185]{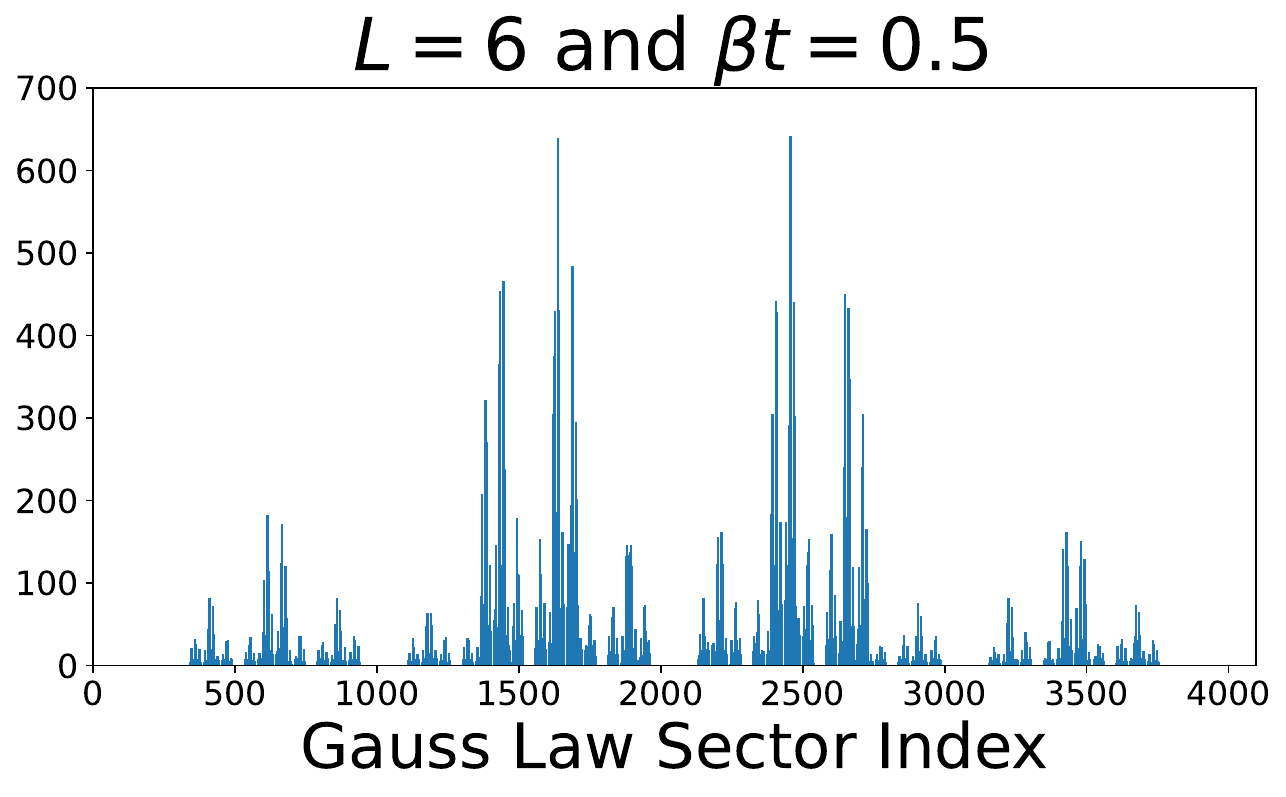}
    \includegraphics[scale=0.185]{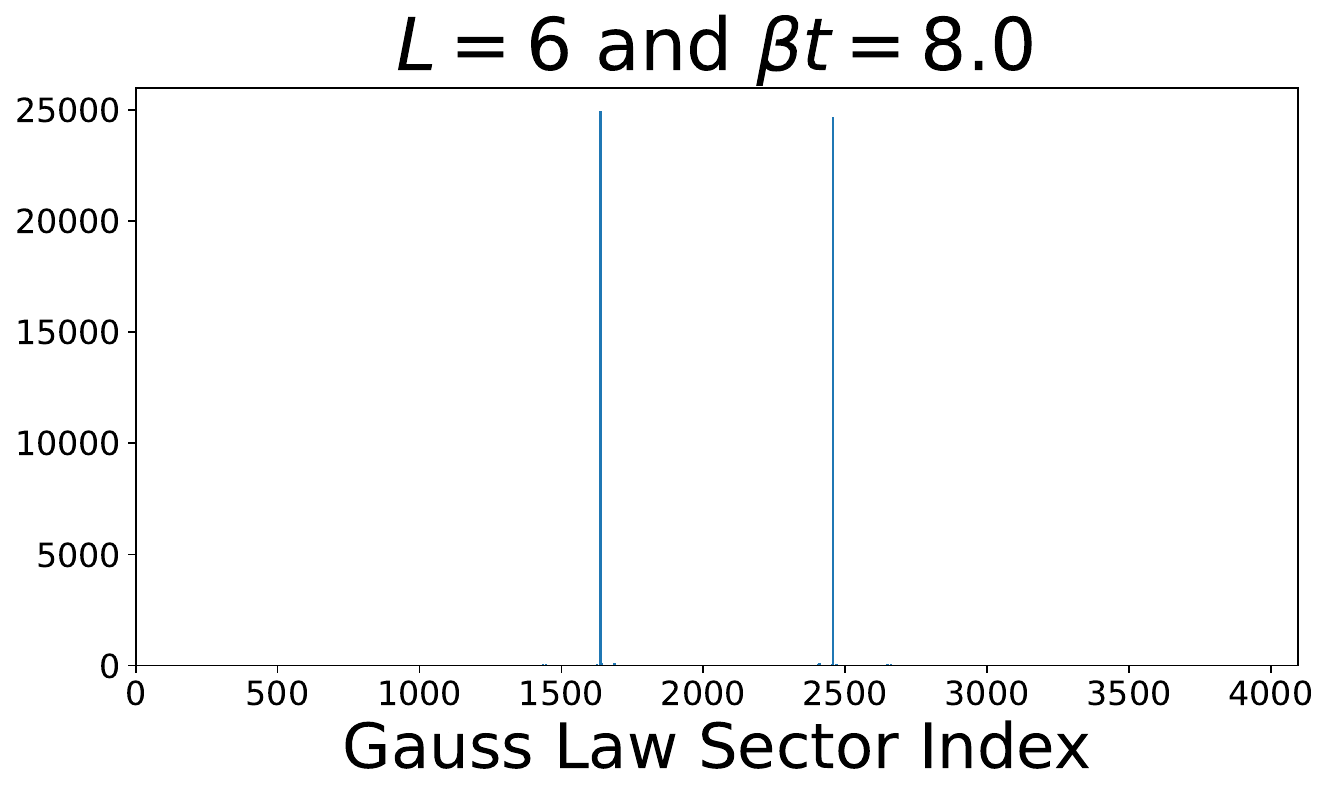}
    \includegraphics[scale=0.18]{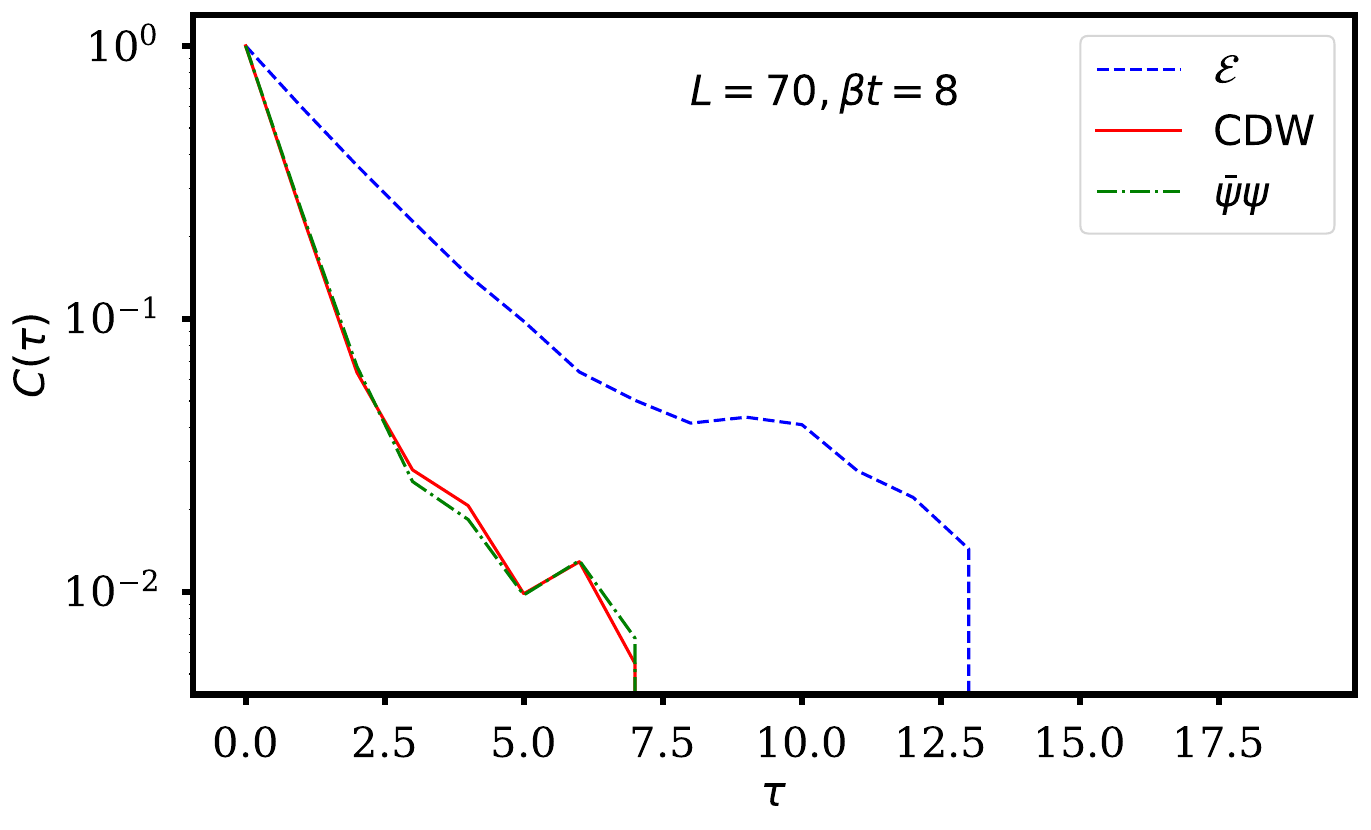}
    \includegraphics[scale=0.18]{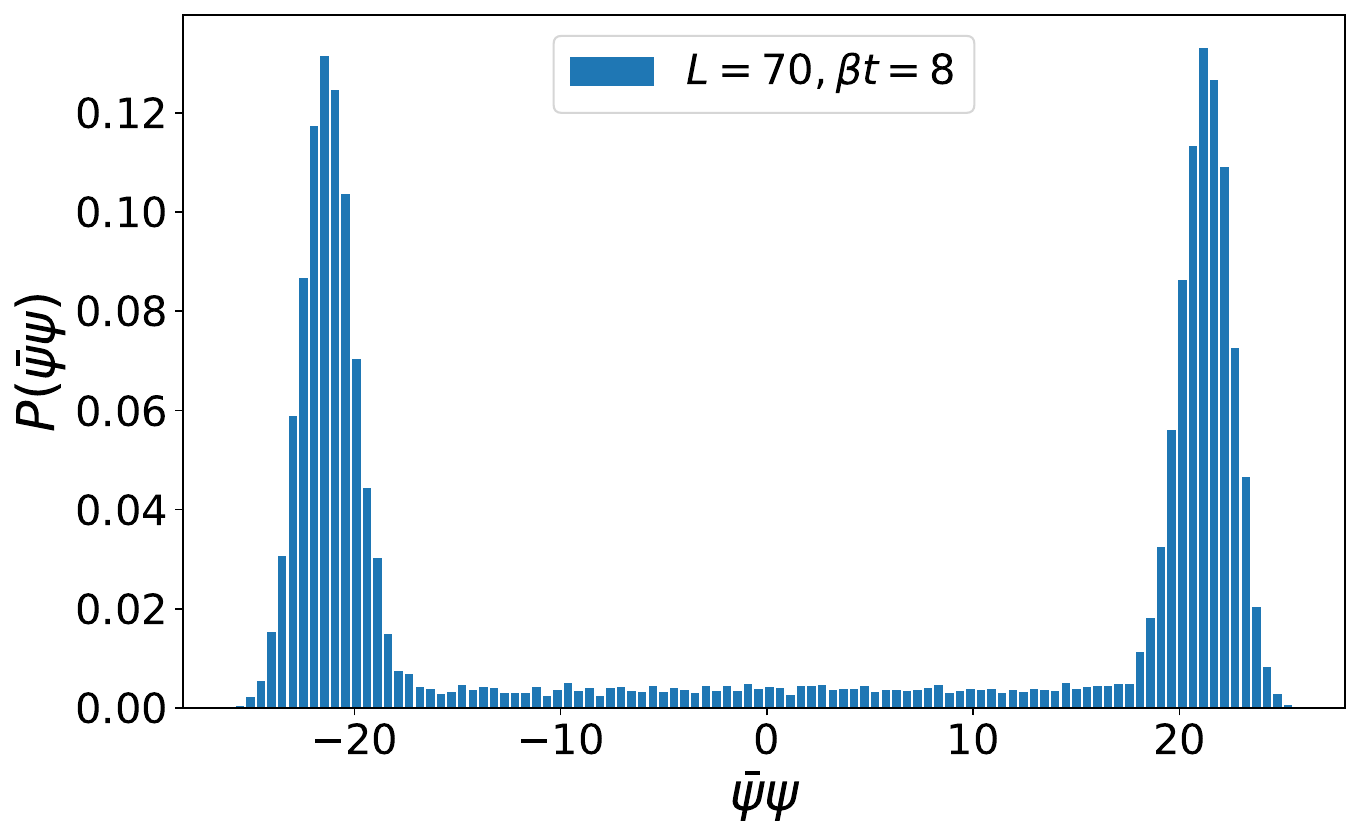}
    \caption{\textit{Clockwise from top left:} (i) and (ii) Number of configurations versus Gauss law sector index $\sum_x\left[G_x+2\right] \cdot 4^x$ 
    (not all indices correspond to actual sectors) for $50000$ equilibrated configurations. Two sectors 
    emerge at large $\beta$: $G_x = 0$ and $G_x = (-1)^x$. (iii) The probability distribution of $\bar{\psi} \psi$, with peaks from the
    two emergent Gauss' Law sectors, indicating that the algorithm efficiently samples all sectors. (iv) The autocorrelation functions for different operators.  }
    \label{fig:GaussLaw}
\end{figure}

\paragraph{Numerical Results.}--
  To illustrate the efficacy of the algorithm, we discuss results obtained by simulating the 
$(1+1)$-d $H^{U(1)}_{N_f=1}$ model in \cref{eq:modelfam}, which is related to the massless 
quantum-link Schwinger model \cite{Banerjee:2012pg, Huang2019} and the PXP model 
\cite{Fendley2004, Surace2020}, where quantum scars were first demonstrated experimentally 
\cite{Bernien_2017}. We simulate the model for different temperatures $\beta=1/T$, 
\emph{without imposing Gauss' law}. A filter may then be applied to study the physics in the 
desired Gauss law sector. The one-dimensional nature of the problem forbids the presence of 
merons, providing a technical simplification. The first non-trivial result is the emergence 
of \emph{two} Gauss' law sectors at low temperatures, as shown in \cref{fig:GaussLaw}. 
For the $\mathbb{Z}_2$ theory in $(1+1)$-d, this result was also observed in \cite{Frank_2020}. 

Generating different Gauss' Law sectors has the benefit that the physics in each sector can be 
easily studied by applying a filter. At low temperatures, this is an ${\cal O}(1)$ effort, but 
becomes exponentially difficult at higher temperatures, since exponentially many sectors will be 
populated. Hence, we note that the efficiency of this meron algorithm for true gauge theories 
(where Gauss' law is imposed) is more suited to the study of quantum phase transitions rather 
than finite-temperature ones. For theories where multiple non-trivial Gauss' Law sectors emerge 
at low temperature, it is possible to study the physics in all sectors without any extra effort.

  Similar to the well-studied Schwinger model \cite{Coleman:1975pw, Coleman:1976uz,Melnikov:2000cc, 
Byrnes:2002nv, Buyens:2016ecr, Banuls:2015sta,Banuls:2016lkq}, our model has the following discrete 
global symmetries: $\mathbb{Z}_2$ chiral symmetry, charge conjugation, $C$, and parity, $P$, \cite{Banerjee:2012pg}, 
whose breaking depends on the strength of the four Fermi coupling. The order parameter sensitive to 
the $P$ or the $C$ symmetry is the total electric flux, ${\cal E} = \frac{1}{L_t} \sum_{x,t} s^3_{x,x+1}$,
while the one for $\mathbb{Z}_2$ chiral symmetry is the chiral condensate, 
$\bar{\psi} \psi = \sum_{x} (-1)^x n_x$. In \cref{fig:GaussLaw} 
we show the probability distribution for $\bar{\psi} \psi$, which samples the two vacua very well, 
indicating that at $T=0$ the symmetry is spontaneously broken. We use these operators to check the algorithm 
against exact diagonalization results, as well as explore other features of the phase at low temperatures. 
We leave these discussions to the Supplementary Material. 
Here we concentrate on the meron algorithm's performance measured via the autocorrelation function:
\begin{equation}
    C_{\cal O} (\tau) = \frac{\braket{ ({\cal O}(i) - \overline{\cal O})({\cal O}(i+\tau) - \overline{\cal O})} }
    {\braket{({\cal O}(i) - \overline{\cal O})^2}},
\end{equation}
where ${\cal O} (i)$ is the measured value at the $i$-th step of the appropriate operator 
(whose average is $\overline{\cal O}$), and is the running index summed over the MC data, 
while the autocorrelations are measured $\tau$ steps apart.
\cref{fig:GaussLaw} shows the $C_{\cal O} (\tau)$ for three different operators: ${\cal E}$, $\bar{\psi} \psi$, 
and ${\rm CDW}$. We note that the bosonic ${\cal E}$ relaxes the slowest, while the 
fermionic operators relax faster. Even for the slowest relaxing mode, the autocorrelation decreases 
by more than an order of magnitude within 10 MC steps for the largest lattice at the lowest temperature, 
demonstrating the efficiency of the algorithm. Finally, we also show the behaviour of the normalized 
susceptibilities corresponding to ${\cal E}$ and the CDW operator as a function of temperature for 
smaller lattices up to $L=22$ in the $G_x=0$ sector. We are able to capture the finite temperature 
crossover.

\begin{figure}
    \includegraphics[scale=0.3]{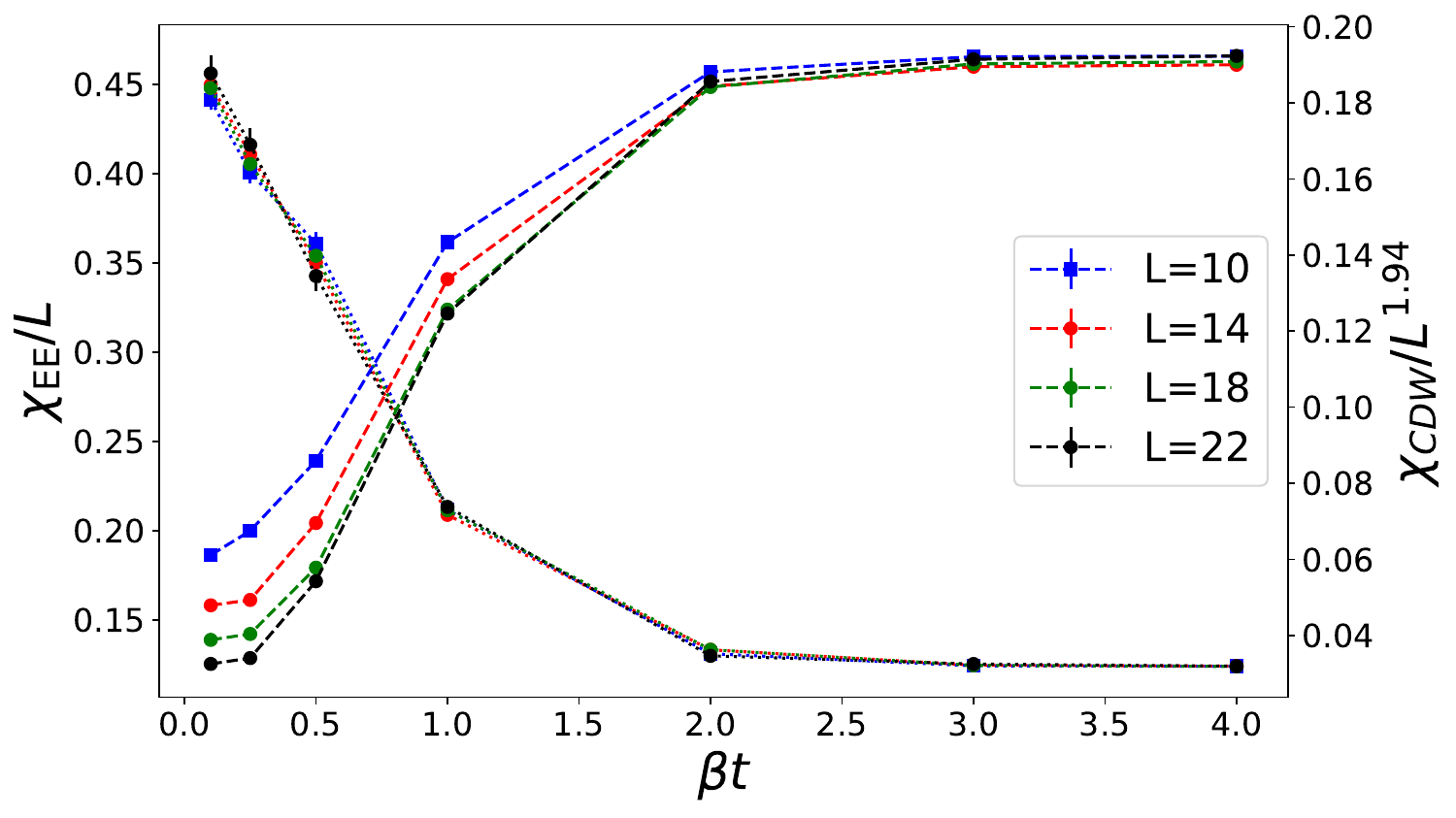}
    \caption{Finite temperature data for $U(1)$ theory in $1+1$-$d$. The dotted lines show the $\chi_{EE}$, 
    which is the susceptibility corresponding to ${\cal E}$. This value rapidly converges to 0.125. On the 
    other hand, the dashed lines trace the $\chi_{CDW}$ which display more finite size effects. 
    Thermal behaviour of both observables indicate that the transition from low to high temperature 
    is a smooth crossover.}
    \label{fig:transition}
\end{figure}

\paragraph{Conclusions.}--
 We have generalized the construction of the meron algorithm to cases where staggered fermions are
coupled to quantum link gauge fields. This construction of the Monte Carlo algorithm is agnostic to the space-time 
dimension, and paves the way for ab-initio studies of large scale gauge-fermionic system with odd or even numbers 
of fermionic flavors, and includes models not simulable using DQMC. While we are able to simulate low temperatures 
at fixed values of gauge coupling by using two breakups, $A$ and $D$, it is possible to add different microscopic 
terms by increasing the allowed ways of bonding the fermions and gauge links. We have also indicated how to include 
multiple flavors, and multi-flavor interactions. Our investigations open up avenues to study quantum link gauge 
theories coupled to fermions in higher dimensions, which are almost certain to exhibit quantum phase transitions
\cite{Janssen_20201}. Since the physics of Abelian gauge fields represented 
by half-integer spins are sometimes related to quantum field theories at $\theta = \pi$ \cite{Zache:2021ggw}, where 
$\theta$ is the topological angle, our numerical method also promises to increase our knowledge of quantum field 
theories with non-trivial topologies. Possible future extensions include gauge fields with larger spin representation 
and non-Abelian gauge fields as well. Our methods can be extended to gauge fields with larger spin representation, and 
hopefully to non-Abelian gauge fields as well, to tackle realistic interacting systems of increasing complexity in 
particle and condensed matter physics.

\paragraph{Acknowledgments.}-- We would like to thank Joao Pinto Barros, Shailesh Chandrasekharan and Uwe-Jens Wiese 
for illuminating discussions and helpful comments. Research of EH at the Perimeter Institute is supported in part 
by the Government of Canada through the Department of Innovation, Science and Economic Development and by the 
Province of Ontario through the Ministry of Colleges and Universities. This work used Bridges 2 at the Pittsburgh 
Supercomputing Center through allocation PHY170036 from the Advanced Cyberinfrastructure Coordination Ecosystem: 
Services \& Support (ACCESS) program, which is supported 
by National Science Foundation grants \#2138259, \#2138286, \#2138307, \#2137603, and \#2138296.
D.B. acknowledges assistance from SERB Starting Grant No. SRG/2021/000396-C from the DST (Government of India).

\bibliography{refs}
\bibliographystyle{unsrt}

\clearpage
\newpage
\appendix

\section{Supplementary Material}

\subsection{$\mathbb{Z}_2$-Symmetric Algorithm}
The breakups for the $\mathbb{Z}_2$-symmetric model defined in \cref{eq:modelfam} and \cref{eq:local} 
are given in Table \ref{tab:z2}. Their weights must satisfy the same equations as for the $U(1)$ theory,
\begin{equation}
      \begin{aligned}
    w_A &=1\\
    w_D &= \exp \left(\epsilon t \right) \sinh \epsilon t \\
    w_A + w_D &= \exp\left(\epsilon t \right) \cosh \epsilon t ,
    \end{aligned}
\end{equation}
which are consistent and solved. The key difference is that more plaquettes are allowed for the $\mathbb{Z}_2$ 
theory compared to the $U(1)$ theory, as seen in the \cref{tab:z2}.

The algorithm then proceeds the same way described in the main text. Note that there are the same number and types
of breakups, and their weights are the same as for a simulation of the $t$-$V$ model. The key difference for the 
gauge theory is that the $A$-breakup involves an additional vertical spin line. This causes clusters rather than 
loops to be formed, which constrain the types of fermionic worldlines that are allowed compared to the loops.

\begin{center}
\begin{table}
    \centering
    \begin{tabular}{ m{2.7cm} | m{2.9cm}}
    \hline
         Plaquettes & Breakups  \\
         \hline
         \includegraphics[width=1.3cm]{Figures/pq1.pdf} \includegraphics[width=1.3cm]{Figures/pq2.pdf}   &  
         \includegraphics[width=1.3cm]{Figures/pa.pdf}
         \includegraphics[width=1.3cm]{Figures/blank.pdf}
         \\
         \hline
         \includegraphics[width=1.3cm]{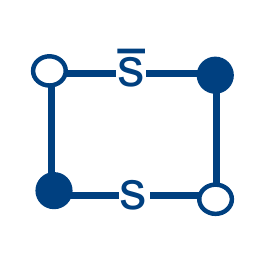}\includegraphics[width=1.3cm]{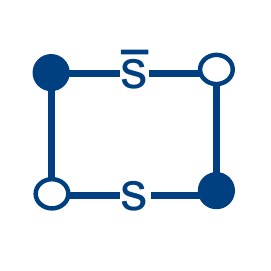} &
         \vspace{.1cm}
         \includegraphics[width=1.3cm]{Figures/dbreakup2.pdf} \\
         \hline
         \includegraphics[width=1.3cm]{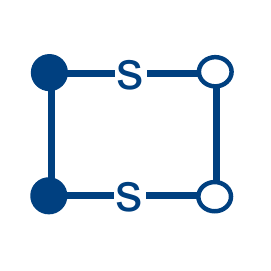}\includegraphics[width=1.3cm]{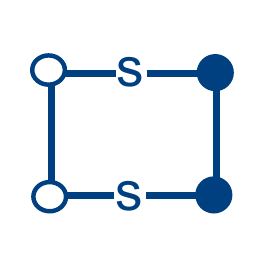} &
         \includegraphics[width=1.3cm]{Figures/pa.pdf}  \includegraphics[width=1.3cm]{Figures/dbreakup2.pdf}    
         \vspace{.05cm}\\
         \hline
    \end{tabular}
    \caption{Plaquettes and breakups for the $\mathbb{Z}_2$-symmetric Hamiltonian.}
    \label{tab:z2}
\end{table}
\end{center}

\subsection{More than One Layer of Fermions and Spins}
As mentioned in the main text of the  paper, the models in (\cref{eq:modelfam}) are sign-problem-free and 
the meron cluster method applies to them. The weights of the matrix elements for one layer of the $U(1)$ 
model were shown explicitly. Here we give the weights for the $N_f$-layer model for the $U(1)$ theory and 
$\mathbb{Z}_2$ theory explicitly so that it is clear that they also are sign-problem-free and amenable to 
the meron cluster method.

First, for $N_f=1$ of the discrete-time version $U(1)$ theory, we can write out the matrix elements for a 
single bond as
\begin{equation}
\begin{aligned}
   & e^{\epsilon t \left(\left[\sigma^1 + \mathbbm{1}\right]_2 \oplus \left[0\right]_6\right)} \\
   &\qquad= \left[\left(\frac{1}{2}e^{2\epsilon t} + \frac{1}{2}\right)\mathbbm{1} +
   \left(\frac{1}{2}e^{2\epsilon t} - \frac{1}{2}\right)\sigma^1\right]_2 \oplus \mathbbm{1}_6.
    \end{aligned}
\end{equation}
These expressions for the diagonal and off-diagonal matrix elements in the $2\times 2$ space are 
the $e^{\epsilon t} \cosh(\epsilon t)$ and $e^{\epsilon t} \sinh(\epsilon t)$ from the main text, and because
\begin{equation}
    1 + \left(\frac{1}{2}e^{2\epsilon t} - \frac{1}{2}\right) = \left(\frac{1}{2}e^{2\epsilon t} + \frac{1}{2}\right)
\end{equation}
we are able to use the $A$ and $D$ breakups to simulate the system.

This mathematical form for the weights is more obviously generalizable to the $N_f$-flavor version of the 
$U(1)$ theory, where the matrix elements for a bond are
\begin{equation}
  \exp\left\{\epsilon t^{N_f} \left(\left[\left(\sigma^1 + \mathbbm{1}\right)^{\otimes N_f}\right]_{m} 
  \oplus \left[0\right]_{n}\right)\right\} ,
\end{equation}
where $m = 2^{N_f}$ and $n = 2^{3N_f} - 2^{N_f}$, performing the exponentiation then yields
\begin{equation}
\begin{aligned}
    & \left[\left(\frac{1}{2^{N_f}} e^{\epsilon (2t)^{N_f}} + \left(1 - \frac{1}{2^{N_f}}\right)\right)\mathbbm{1}_m \right. \\
    &\left. + \left(\frac{1}{2^{N_f}} e^{\epsilon (2t)^{N_f}} - \frac{1}{2^{N_f}}\right) O^{U(1)}_m\right] \oplus \mathbbm{1}_n ,
    \end{aligned}
\end{equation}
where we define $O^{U(1)}_m$ to be an ``off-diagonal'' matrix for the $U(1)$ theory, which is $m\times m$ 
and has entries of $1$ everywhere except for on the diagonal, where the entries are zero.

As required all weights are positive, and we have
\begin{equation}
\begin{aligned}
   & 1 + \left(\frac{1}{2^{N_f}} e^{\epsilon (2t)^{N_f}} - \frac{1}{2^{N_f}}\right)\\
    &= \left(\frac{1}{2^{N_f}} e^{\epsilon (2t)^{N_f}} + \left(1 - \frac{1}{2^{N_f}}\right)\right),
    \end{aligned}
\end{equation}
allowing us to use $A^{\otimes N_f}$ and $D^{\otimes N_f}$ breakups to simulate the system.

The $\mathbbm{Z}_2$ matrix elements are found from
\begin{equation}
\begin{aligned}
 & \exp\big\{\epsilon t^{N_f} \left(\left[\left(\left(\sigma^1 + \mathbbm{1}\right)\oplus\left(\sigma^1 + \mathbbm{1}\right) \right)^{\otimes N_f}\right]_{m} \right.  \\
& \qquad\quad\qquad\qquad\qquad\qquad\qquad\qquad\qquad\left.\oplus \left[0\right]_{n}\right) \big\},
  \end{aligned}
\end{equation}
where this time $m= 2^{2N_f}$ and $n = 2^{3N_f}- 2^{2N_f}$. Performing the exponentiation then yields
\begin{equation}
\begin{aligned}
   & \left[\left(\frac{1}{2^{N_f}} e^{\epsilon (2t)^{N_f}} + \left(1 - \frac{1}{2^{N_f}}\right)\right)\mathbbm{1}_m \right. \\
    &\left. + \left(\frac{1}{2^{N_f}} e^{\epsilon (2t)^{N_f}} - \frac{1}{2^{N_f}}\right) O_m^{\mathbb{Z}_2}\right] \oplus \mathbbm{1}_n ,
    \end{aligned}
\end{equation}
where the off-diagonal $O_m^{\mathbb{Z}_2}$ matrix is defined as
\begin{equation}
    O_m^{\mathbb{Z}_2} = \left[\left(\sigma^1+1\right)\oplus \left(\sigma^1 + 1\right)\right]^{\otimes N_f} - \mathbbm{1}_m,
\end{equation}

\begin{figure*}
    \centering
    \includegraphics[width=3.5in]{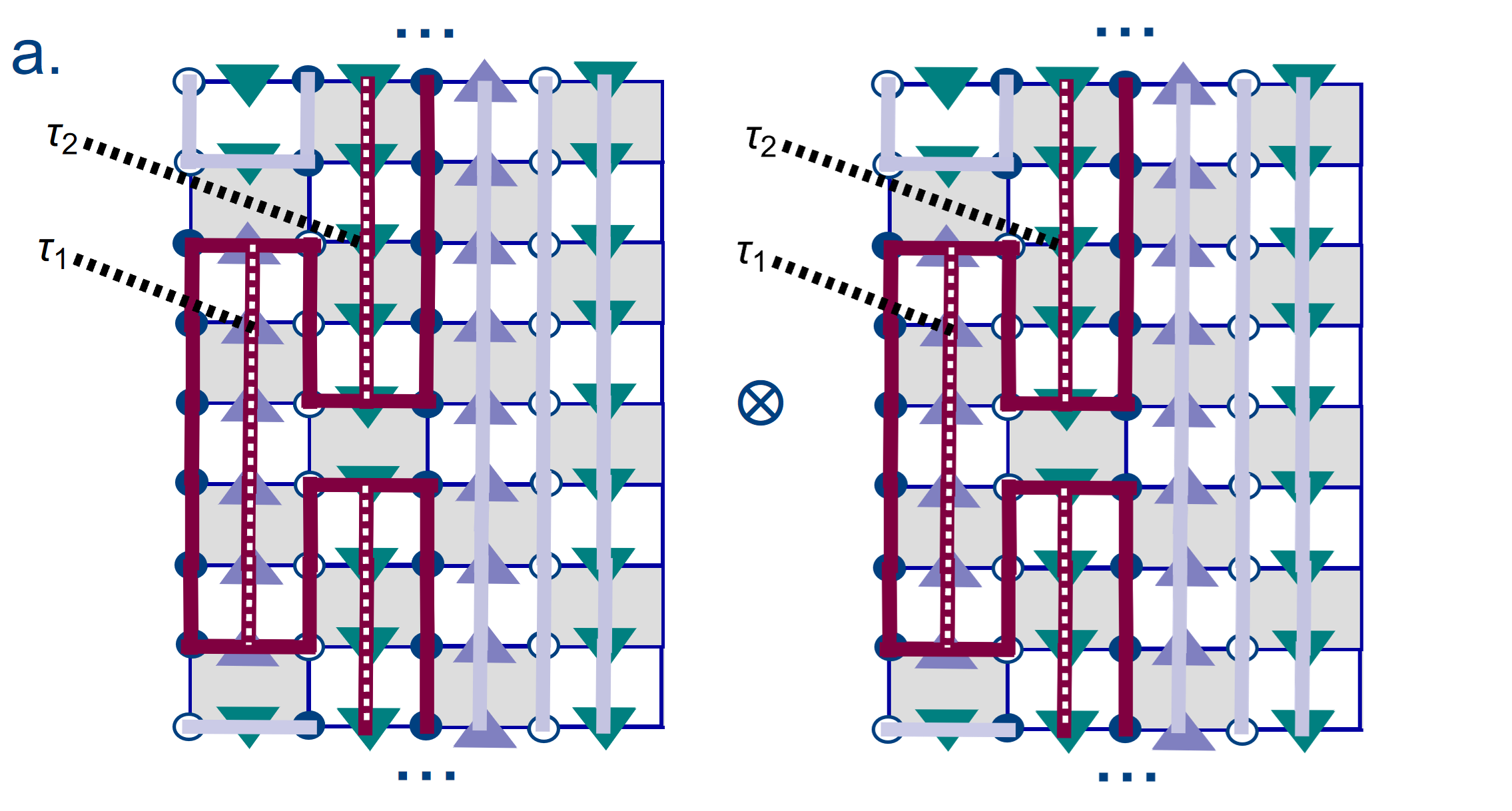}
    \includegraphics[width=3.5in]{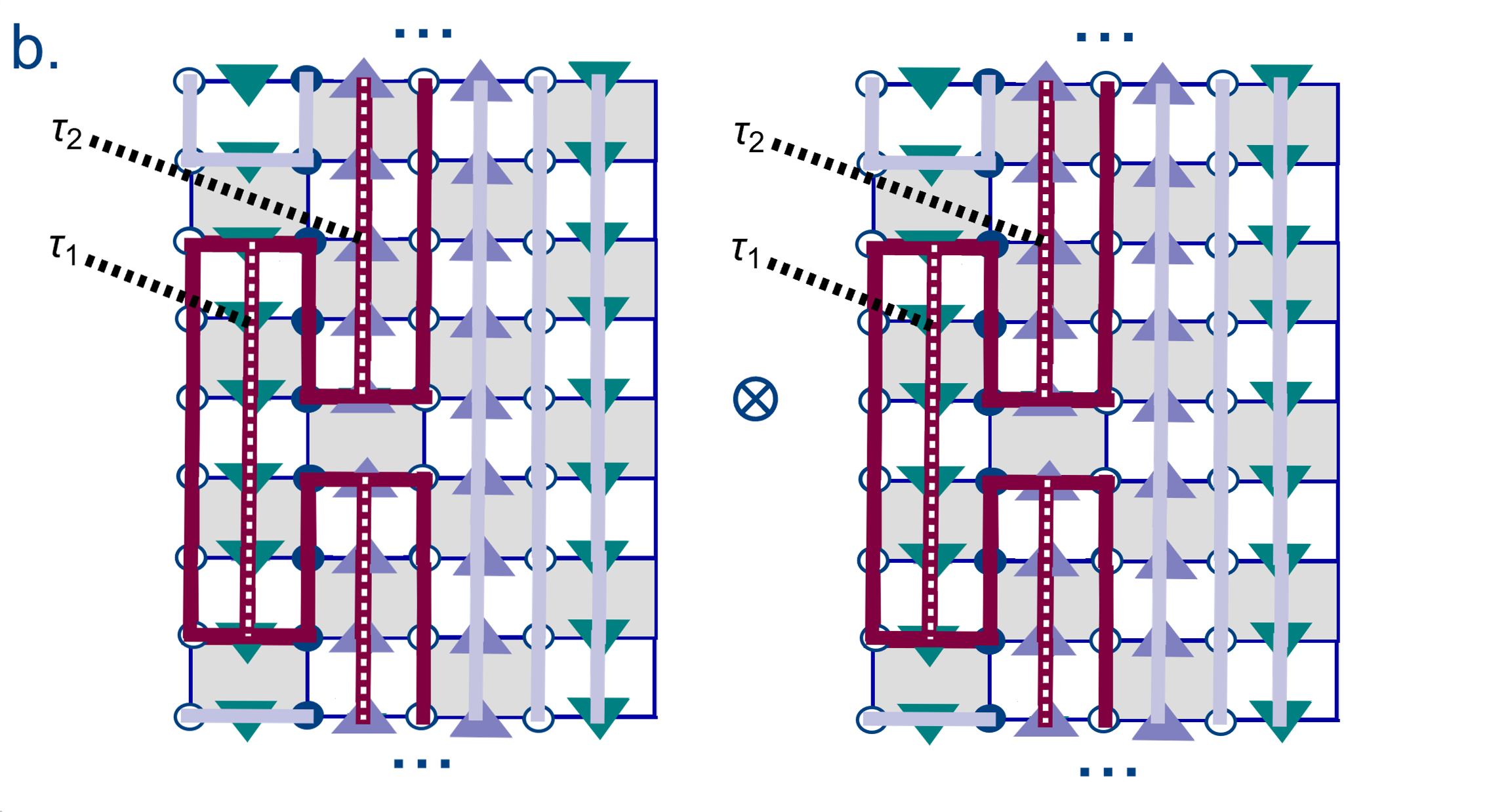}
    \includegraphics[width=3.5in]{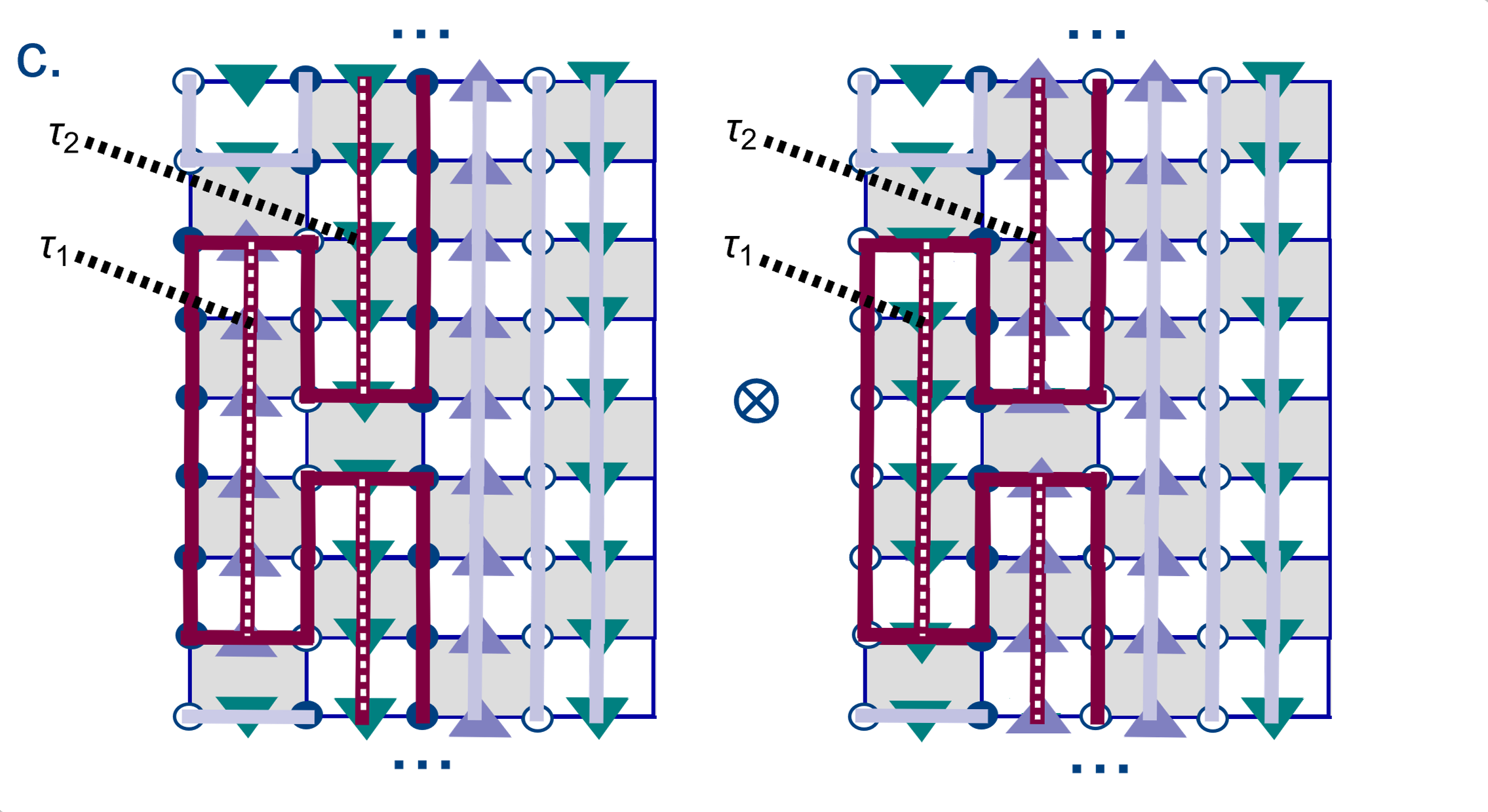}
    \includegraphics[width=3.5in]{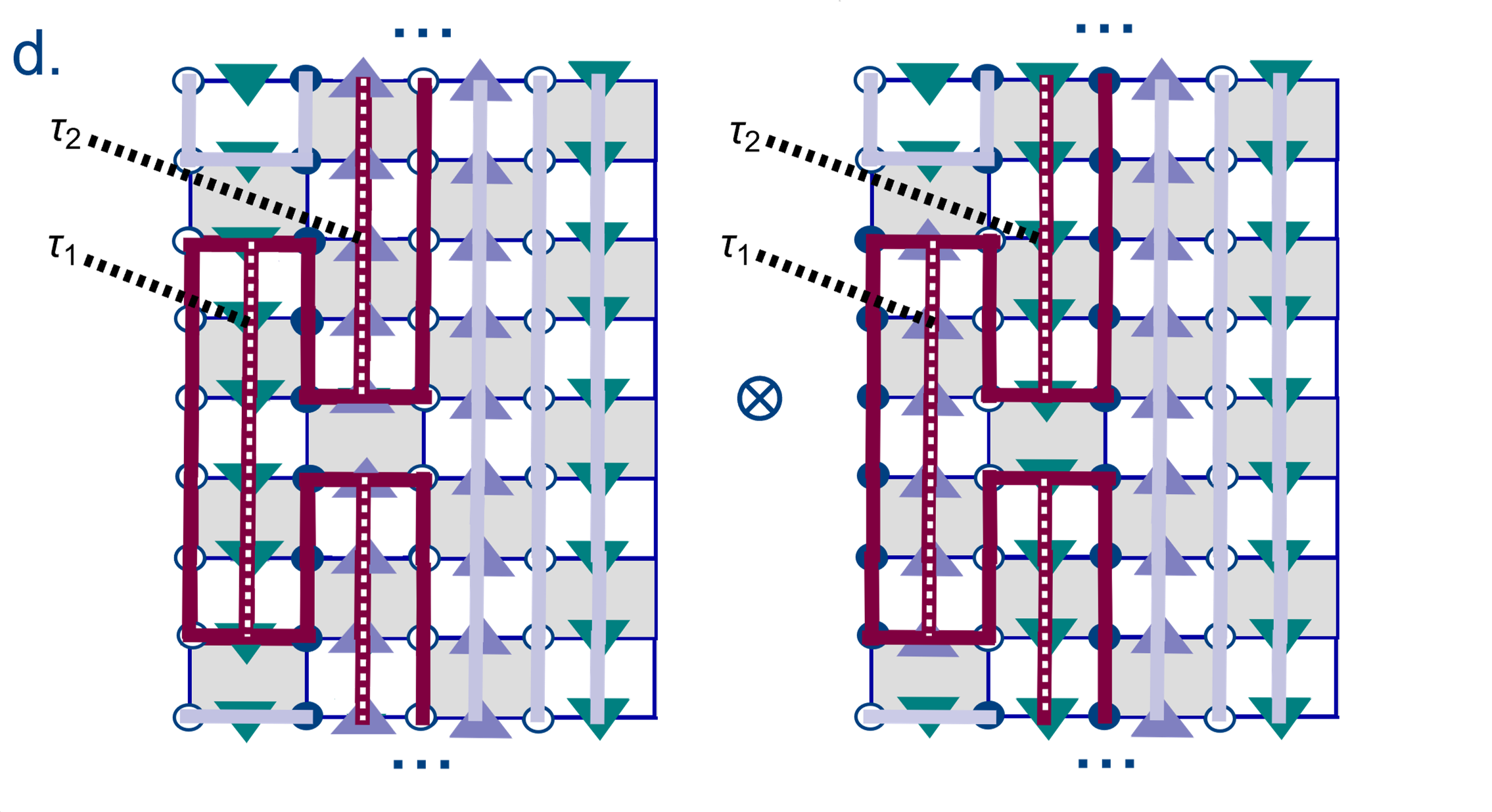}
    \caption{The spin factors of the weights for the portion of these terms that are shown 
    (assuming the burgundy cluster is a meron) are: $e^{\tau_1 J /4} e^{\tau_2 J / 4}$ for 
    (a) and (b), and $e^{-\tau_1 J /4} e^{-\tau_2 J / 4}$ for (c) and (d). Because the merons 
    in (a) and (b) are positive and the merons in (c) and (d) are negative, the fully summed 
    weight of the meron configuration will be positive. This assumes a staggered reference 
    configuration where both layers are identical.}
    \label{fig:2layers}
\end{figure*}

\subsection{Additional Field Terms for the $U(1)$ Model}
As discussed in the main text, the $U(1)$ model families have additional diagonal terms for the spins 
that may be added:
\begin{equation}
    H^{U(1)}_{N_f=2}\rightarrow H^{U(1)}_{N_f=2} + J\sum_{\left\langle xy\right\rangle} s^3_{xy,1} s^3_{xy,2},
\end{equation}
Here we discuss why these additions are sign-problem-free. Note that we will focus on the discrete-time 
version of the Meron Cluster algorithm, but the result also straightforwardly carries to continuous time. 
This addition is sign-problem-free for a similar reason that the addition of the Hubbard-$U$ term is 
sign-problem-free, as discussed in \cite{Liu:2020ygc}.

We apply the discrete time partition function expansion defined in the main text with general $d$ spatial 
dimensions, which is given for the new Hamiltonian $H'$ by
\begin{equation}
\begin{aligned}
    &{\cal Z} ={\mathrm{Tr}} \left(e^{-\beta H'}\right)\\
    &\; = \sum_{\left\{s,n\right\}} \left\langle s_1,n_1\right| e^{-\epsilon H_{a_T}'}\left|s_{T N_t},n_{T N_t}\right\rangle\\ &\qquad \left\langle s_{T N_t},n_{T N_t}\right| e^{-\epsilon H_{a_{T-1}}'}\left|s_{T N_t-1},n_{T N_t-1}\right\rangle...\\
   &\qquad\left\langle s_3,n_3\right| e^{-\epsilon H_{a_{2}}'}\left|s_2,n_2\right\rangle\left\langle s_2,n_2\right|e^{-\epsilon H_{a_1}'}\left|s_1,n_1\right\rangle .
    \end{aligned}
\end{equation}
This Hamiltonian is Trotterized into $T$ sets of operators where the operators within each set commute 
with each other, $H' = H_{a_1}' + H_{a_1}'+...+H_{a_T}'$, thus there are $T\cdot N_t$ timeslices in this 
formula. In one spatial dimension, as seen in the main text, $T$ is $2$ and the Hamiltonian was divided 
into $H_e$ and $H_o$, for even and odd, respectively.

Writing in terms of the original $U(1)$ Hamiltonian $H$ (as simulated in this paper), we have
\begin{equation}
\begin{aligned}
    &{\cal Z} ={\mathrm{Tr}} \left(e^{-\beta H}\right)\\
    &\; = \sum_{\left\{s,n\right\}} \left\langle s_1,n_1\right| e^{-\epsilon \left(H_{a_T} + H^s_{a_T}\right)}\left|s_{T N_t},n_{T N_t}\right\rangle\\ &\left\langle s_{T N_t},n_{T N_t}\right| e^{-\epsilon(H_{a_{T-1}} + H_{a_{T-1}}^s)}\left|s_{T N_t - 1},n_{T N_t - 1}\right\rangle\\
    &\qquad\quad...\left\langle s_3,n_3\right|e^{-\epsilon \left(H_{a_2}+ H_{a_2}^s\right)}\left|s_2,n_2\right\rangle\\
    &\quad\qquad\qquad \left\langle s_2,n_2\right|e^{-\epsilon \left(H_{a_1} + H_{a_1}^s\right)}\left|s_1,n_1\right\rangle ,
    \end{aligned}
\end{equation}
where for example in $(1+1)$-$d$ we would have $a_1=o$, $a_2 = e$, and
\begin{equation}
\begin{aligned}
    H_e^s &= \sum_{x} s^3_{2x,2x+1,1} s^3_{2x,2x+1,2},\\ H_o^s &= \sum_{x} s^3_{2x+1,2x+2,1} s^3_{2x+1,2x+2,2}.
    \end{aligned}
\end{equation}
Here we note that because these additional terms respect the gauge symmetry, we have for a generic Trotter 
operator set $a$,
\begin{equation}
\begin{aligned}
    e^{-\epsilon \left(H_a + H_a^s\right)} \left| s_i n_i\right\rangle &=  e^{-\epsilon H_{a} }  e^{-\epsilon H_{a}^s } \left| s_i n_i\right\rangle \\
    &= e^{-\epsilon \left(h_{a}^s (s_i)\right)}  e^{-\epsilon H_a}   \left| s_i n_i\right\rangle,
    \end{aligned}
\end{equation}
where $h_a^s(s_i)$ is just a number that depends on the spin configuration at particular timeslice $i$, 
which is $s_i$. 

Thus, we end up with a partition function very similar to our original one, but with the worldlines 
weighted slightly differently:
\begin{equation}
\begin{aligned}
     &{\cal Z} ={\mathrm{Tr}} \left(e^{-\beta H}\right)\\
    &\; = \sum_{\left\{s,n\right\}} e^{-\epsilon \sum_i\left( h_{a_1}^s(s_{iT+1})+h_{a_2}^s(s_{iT+2})...h_{a_T}^s(s_{iT+T})\right)} \\ & \left\langle s_1,n_1\right| e^{-\epsilon H_{a_T}'}\left|s_{TN_t},n_{TN_t}\right\rangle\left\langle s_{TN_t},n_{TN_t}\right|e^{-\epsilon H_{a_{T-1}}'}\\ &\qquad\qquad ...e^{-\epsilon H_{a_2}'}\left|s_2,n_2\right\rangle\left\langle s_2,n_2\right|e^{-\epsilon H_{a_1}'}\left|s_1,n_1\right\rangle .
    \end{aligned}
\end{equation}
Because of these additional weight factors, the merons no longer have zero weight, so the question is 
instead whether their weight can be guaranteed to be positive.

To figure out the meron weights, we consider a generic meron, a portion of which is illustrated in 
\cref{fig:2layers}, noting that merons only appear in two spatial dimensions and higher, so the images 
in the figure would represent a cross-section of a $(2+1)$-$d$ (or higher dimensional) lattice, with 
two Trotter time slices shown, and the others squashed since they are irrelevant to this portion of the 
meron. Here, we assume a staggered reference configuration of fermions that is identical for both spin 
layers. In the portion highlighted, the weight factors for the highlighted meron portions of the 
configurations (a) and (b) will have a positive sign because both layers have the same fermion 
configurations, and thus the sign contribution from each layer will be identical, leading to a positive 
product. The configurations (c) and (d) will have a negative sign factor coming from the highlighted 
meron because fermion occupations are flipped between the two layers. We have already established that 
the weights coming from the fermionic term are identical. The contributing factors for this meron 
portion that come from the spins are
\begin{equation}
    e^{\tau_1 J / 4} e^{\tau_2 J / 4}
\end{equation}
for configurations (a) and (b), and 
\begin{equation}
    e^{-\tau_1 J / 4} e^{-\tau_2 J /4}
\end{equation}
for configurations (c) and (d), because in the $U(1)$ theory spin degrees of freedom are tied to the 
fermion occupation numbers on either side of the $D$-breakups, and so if the fermions match in the two 
layers the spins must also match, and similarly if the fermions do not match in the two layers for the 
meron the spins also cannot match. Thus, the sum of the four configurations must be a positive number. 
We note that this also can be done for negative $J$-coupling, but there the staggered reference 
configuration changes to be opposite occupation numbers for the two layers.

\subsection{Merons and Autocorrelation in the Gauge Models}
For purely fermionic models, as shown in \cite{Chandrasekharan:1999cm}, a loop is a meron if the 
quantity $n_w + n_h/2$ is even, where $n_w$ is the number of temporal windings and $n_h$ is the 
number of fermionic hops in the loop. For the gauge extension of this algorithm, we now have the 
possibility of binding two or more loops to each other through the gauge, and thus we have the 
following criterion for a cluster to be a meron
\begin{equation}
\mathrm{Meron \; if} \left\{\begin{array}{cc} n_w + n_h/2\; \mathrm{odd}, & \mathrm{even\;} \# \mathrm{\;of \;loops}\\
n_w + n_h/2\; \mathrm{even}, & \mathrm{odd\;} \# \mathrm{\;of \;loops}\end{array}\right.
\label{meroncriterion}
\end{equation}
It can be seen immediately that (\ref{meroncriterion}) reduces to the original definition in the case 
of one loop, and to understand the extension we note that in the case of two loops bound together 
(or more generally an even number of loops bound together), the result will be a meron in the case of 
one loop (or an odd number of loops) being a ``loop meron'' by the original loop meron definition, and 
the rest of the loops being non-``loop merons.'' This results in the total $n_w + n_h/2$ being an odd 
number of even numbers plus an odd number of odd numbers, which is an odd number. A similar argument 
follows for the clusters that consist of an odd number of loops.

As mentioned in the main text, in one dimension there should be no merons. Figure 
\ref{clusters} plots the $n_h$ and $n_w$ for $100$ equilibrated clusters at low temperature, and 
indeed we see that by the criterion, none of these clusters are merons.

\begin{figure}
\includegraphics[width=6cm]{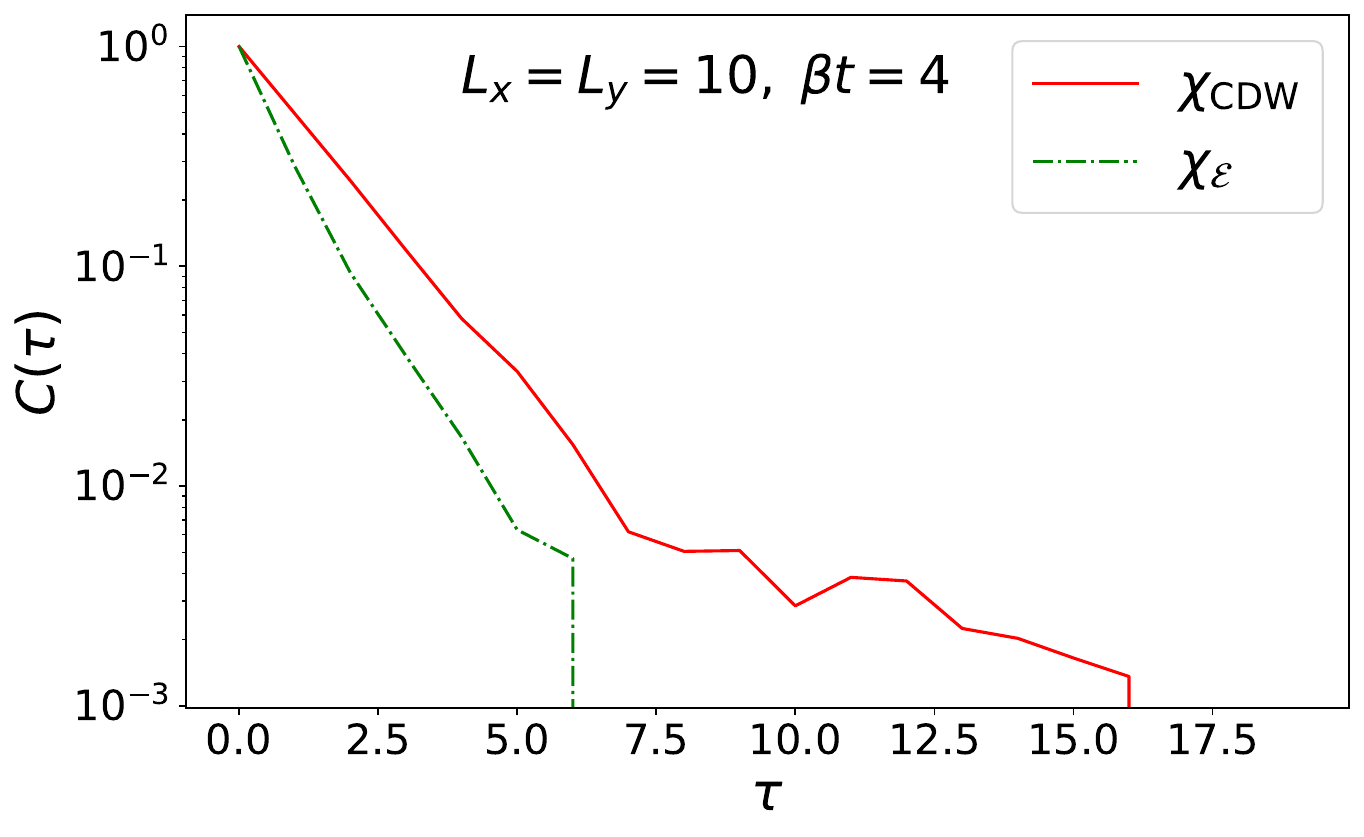}
\caption{The autocorrelation functions for two susceptibility operators in two dimensions.}
\label{autocorr2d}
\end{figure}

\begin{figure}
    \includegraphics[width=6cm]{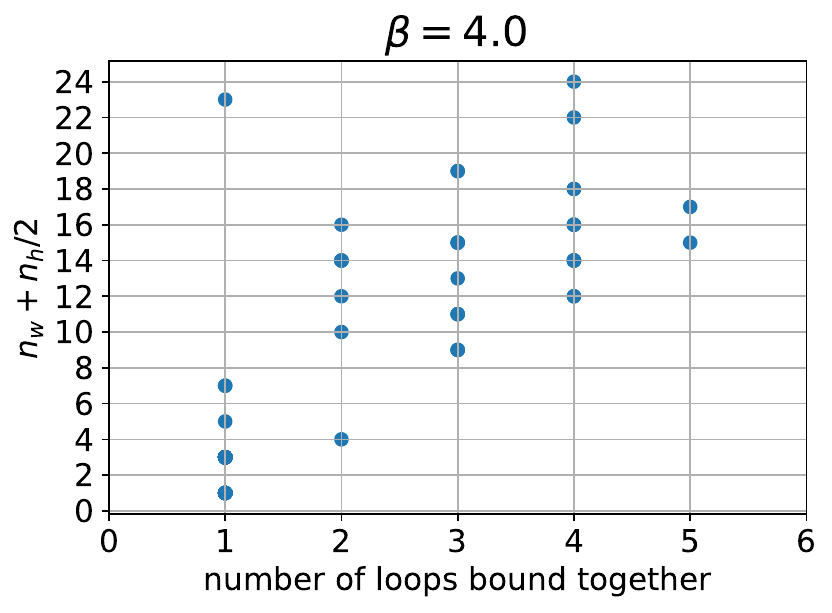}
    \caption{The $n_h$ and $n_w$ for $100$ equilibrated clusters plotted for the $U(1)$ theory in one dimension.}
    \label{clusters}
\end{figure}

In two dimensions, we indeed get merons, and we can compute observables by sampling 
the zero- and two-meron sectors according to the procedure in \cite{Chandrasekharan:1999cm}. 
Figure \ref{autocorr2d} gives a similar autocorrelation calculation to the one given in the main text 
for the $\mathcal{E}$ and CDW susceptibilities on a $10\times10$ lattice in two dimensions, and we see 
that while the autocorrelation times are marginally longer than those in one dimension, they are still 
quite short.

\begin{figure}
    \centering
    \includegraphics[scale=0.35]{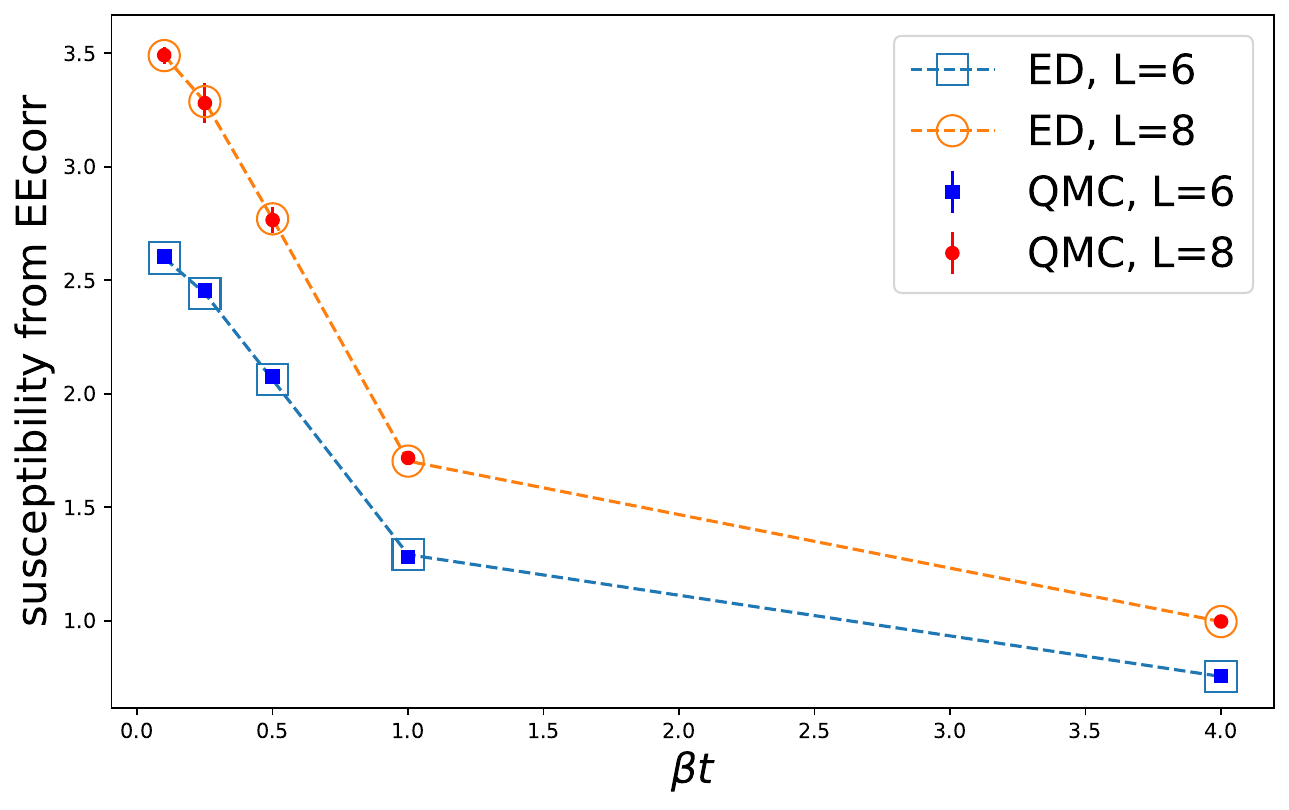} \\
    \includegraphics[scale=0.35]{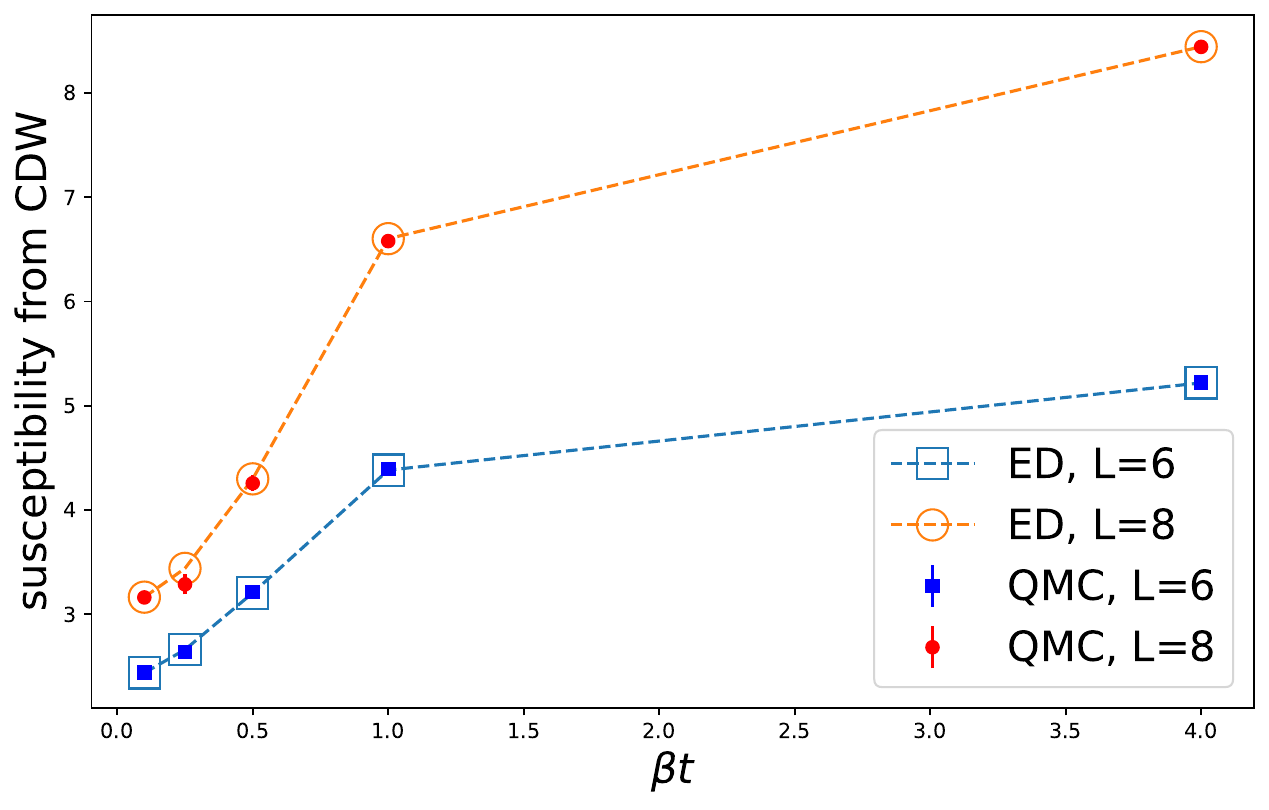} \\
    \includegraphics[scale=0.35]{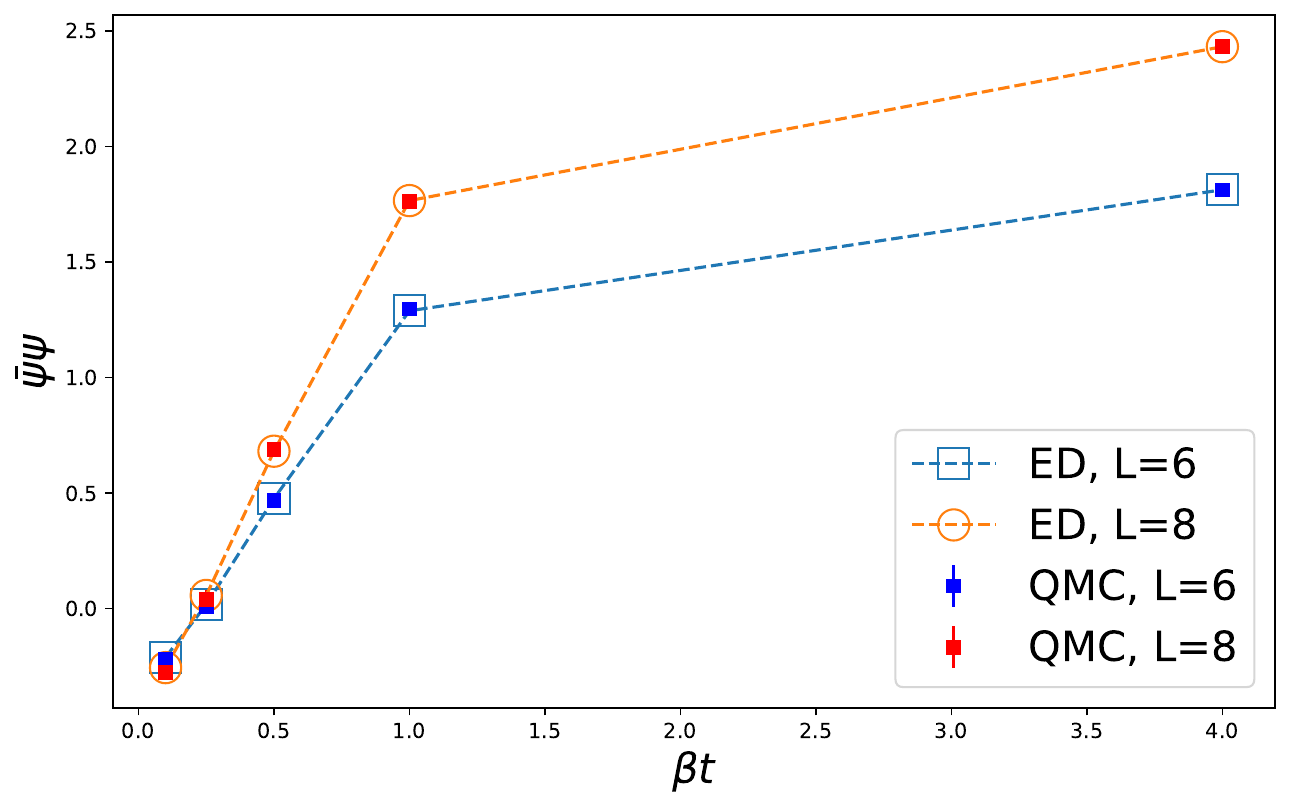}
    \caption{Comparison of Meron Cluster QMC values with the ED values for $\chi_{\mathrm{EEcorr}}$ (top), 
    $\chi_{\mathrm{CDW}}$ (middle), and the bare $\bar{\psi}\psi$ (bottom).}
    \label{fig:compED}
\end{figure}

\begin{table}
\begin{center}
\small
\begin{tabular}{|l||l|l|l|l|l|}
\hline
 $\mathbf{\beta t}$ & $0.1$ & $0.25$ & $0.5$ & $1.0$ & $4.0$  \\
 \hline
\multicolumn{6}{|c|}{$\chi_{\mathrm{EEcorr}}$, $L=6$} \\
\hline
MC & 2.60(3) & 2.46(3) & 2.08(2) &
1.281(7) & 0.7556(7) \\
\hline
ED & 2.598 & 2.442 & 2.062 &
1.291 & 0.7542 \\
\hline
\multicolumn{6}{|c|}{$\chi_{\mathrm{EEcorr}}$, $L=8$} \\
\hline
MC & 3.49(4) & 3.27(9) & 2.76(6) & 1.72(2) & 0.9966(9) \\
\hline
ED & 3.489 & 3.287 & 2.770 & 1.703 & 0.99606 \\
\hline
\multicolumn{6}{|c|}{$\chi_{\mathrm{CDW}}$, $L=6$} \\
\hline
MC & 2.44(4) & 2.64(3) & 3.21(3) & 4.39(1) & 5.2225(20) \\
\hline
ED & 2.438 & 2.661 & 3.205 & 4.383 & 5.2197 \\
\hline
\multicolumn{6}{|c|}{$\chi_{\mathrm{CDW}}$, $L=8$} \\
\hline
MC & 3.16(4) & 3.287(95) & 4.26(8) & 6.58(4) & 8.443(4) \\
\hline
ED & 3.164 & 3.440 & 4.299 & 6.603 & 8.445 \\
\hline
\multicolumn{6}{|c|}{$\bar{\psi}\psi$, $L=6$} \\
\hline
MC & -0.22(2) & 0.017(9) & 0.47(1) & 1.296(6) & 1.8126(8) \\
\hline
ED & -0.210 & 0.0171 & 0.476 & 1.289 & 1.81284 \\
\hline
\multicolumn{6}{|c|}{$\bar{\psi}\psi$, $L=8$} \\
\hline
MC & -0.28(2) & 0.04(4) & 0.68(3) & 1.76(1) & 2.432(1) \\
\hline
ED & -0.255 & 0.057 & 0.681 & 1.766 & 2.4322 \\
\hline
\end{tabular}
\vspace{.5cm}
\caption{Results for the observables $\chi_{\mathrm{EEcorr}}$, $\chi_{CDW}$ and 
$\bar{\psi}\psi$ defined in (\ref{observables}) for the $U(1)$ model in $(1+1)$-$d$ 
calculated with Meron Clusters (MC) and exact diagonalization (ED).}
\label{datatable2}
\end{center}
\end{table}

\subsection{Checks Against ED}
For this paper we implemented the Meron Cluster method for the $U(1)$ theory in $1+1$-$d$, and for small 
lattices ($L=6$ and $L=8$) we have checked the following observables against exact 
diagonalization (ED) calculations:
\begin{equation}
    \begin{aligned}
        \chi_{CDW} &= \frac{1}{L_t} \sum_{x,y,t} (-1)^{x+y} \left(n_x - \frac{1}{2}\right) \left(n_{y}-\frac{1}{2}\right) \\
        \bar{\psi}\psi &= \frac{1}{L_t} \sum_{x,y,t} (-1)^x n_x \\
        \chi_{\mathrm{EEcorr}} &= \frac{1}{L_t} \sum_{x,y,t}  s^3_{x,x+1} \cdot s^3_{y,y+1}.
        \label{observables}
    \end{aligned}
\end{equation}
Table \ref{datatable2} and \cref{fig:compED} summarize these results, which are all for the observables 
in the Gauss law sector $G_x=0$.

\begin{figure}
    \includegraphics[scale=0.18]{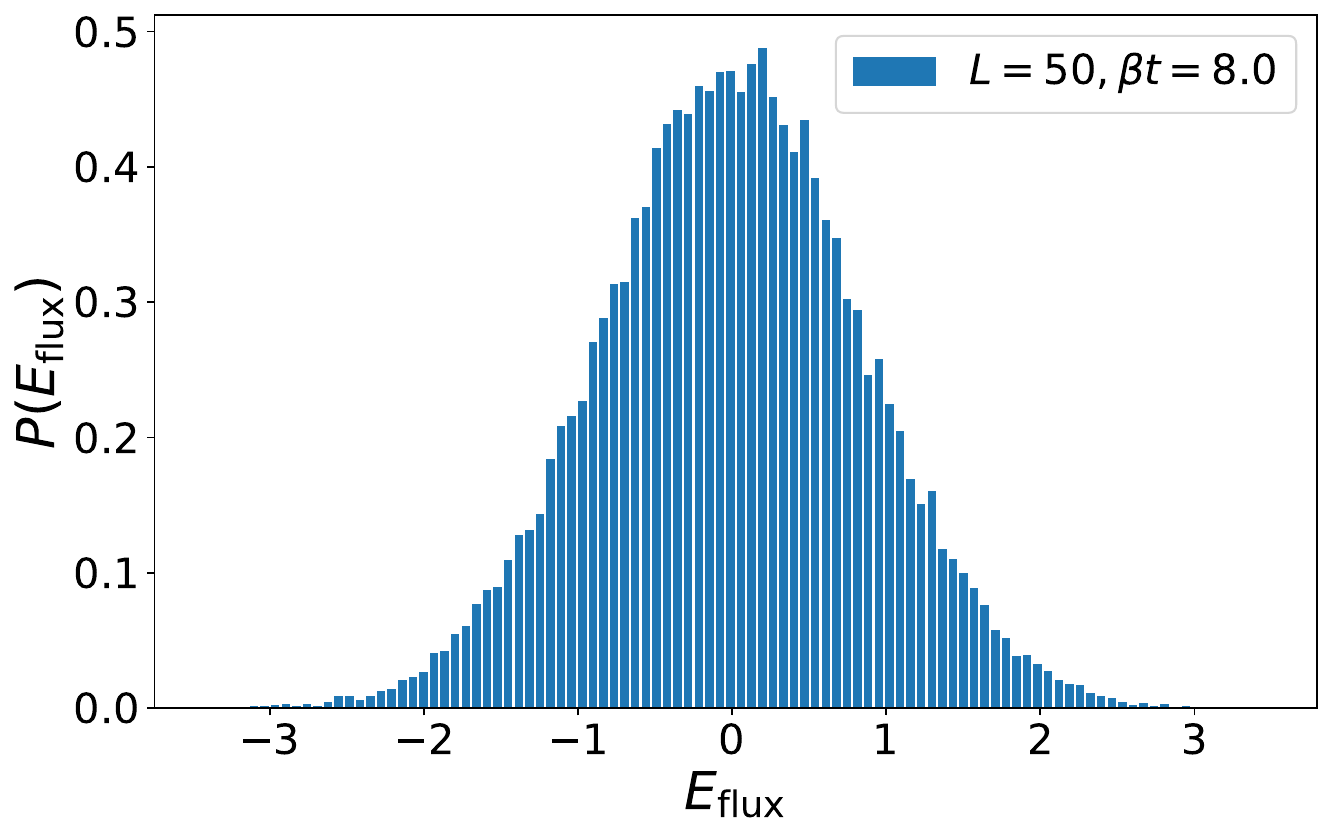}
    \includegraphics[scale=0.18]{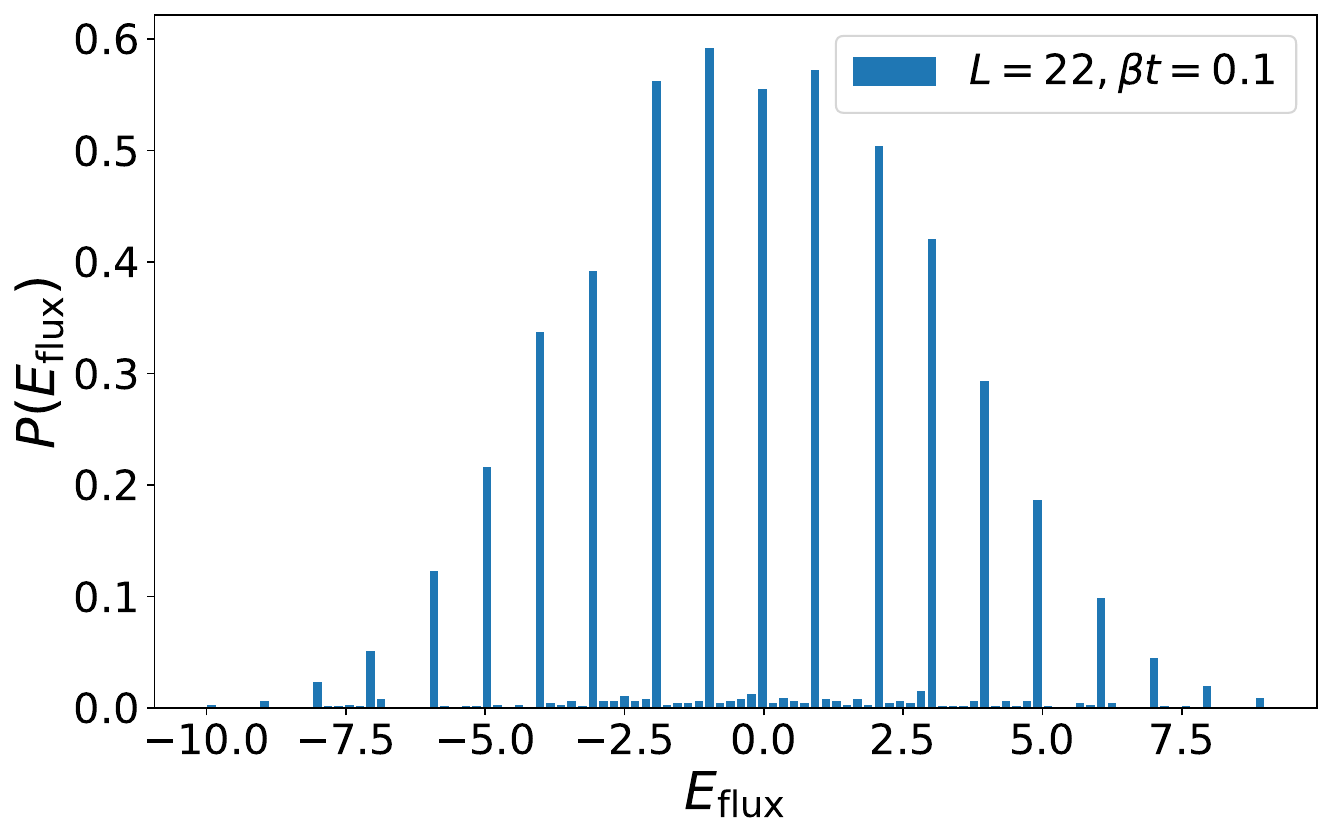}
    \caption{Probability density for the electric flux ${\cal E}$ for low temperature ($L=50$, $\beta t=8.0$) 
    and high temperature ($L=22$, $\beta t=0.1$).}
    \label{fig:EEhist}
\end{figure}

\begin{figure}
    \centering
    \includegraphics[width=3in]{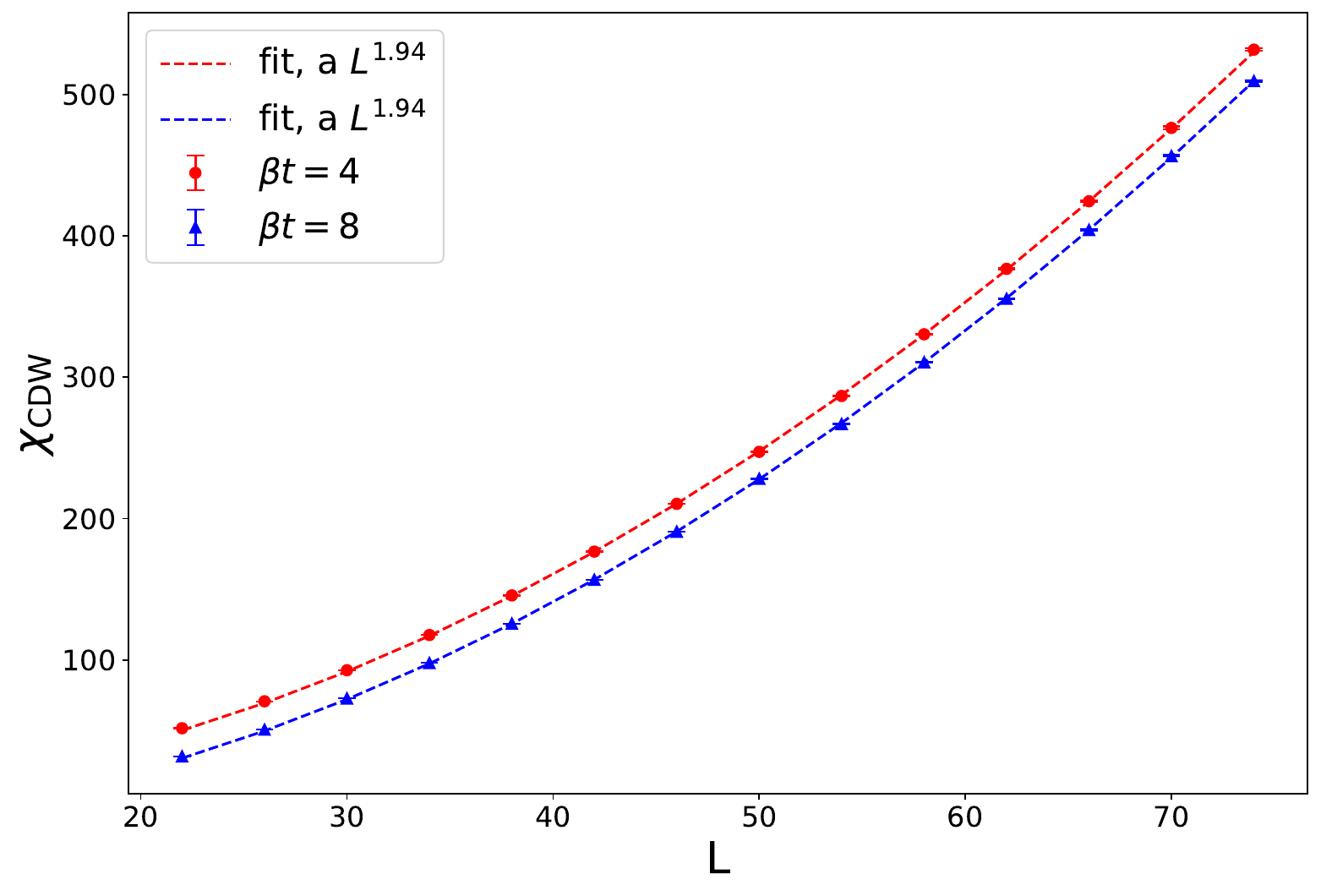}
    \caption{Divergence of the chiral condensate at low temperature.}
    \label{fig:chiDiv}
\end{figure}

\subsection{Large volume physics with the Meron Algorithm}
 Although the fermion gauge coupling in the model we simulate with the meron algorithm is identical to the quantum 
 link Schwinger model with spin-$\frac{1}{2}$, we have additional four fermion terms, which behave differently from 
 the staggered mass term. While this interaction also induces an effective mass term, the mapping to a massive 
 Schwinger model with renormalized parameters is non-trivial. This makes it hard to connect some of the large
 volume quantities to those usually studied in the literature \cite{Coleman:1975pw, Coleman:1976uz, 
 Melnikov:2000cc, Byrnes:2002nv, Buyens:2016ecr, Banuls:2015sta,Banuls:2016lkq}. Nevertheless, we present an
 empirical study of our measured observables. 

\begin{figure}
    \includegraphics[scale=0.35]{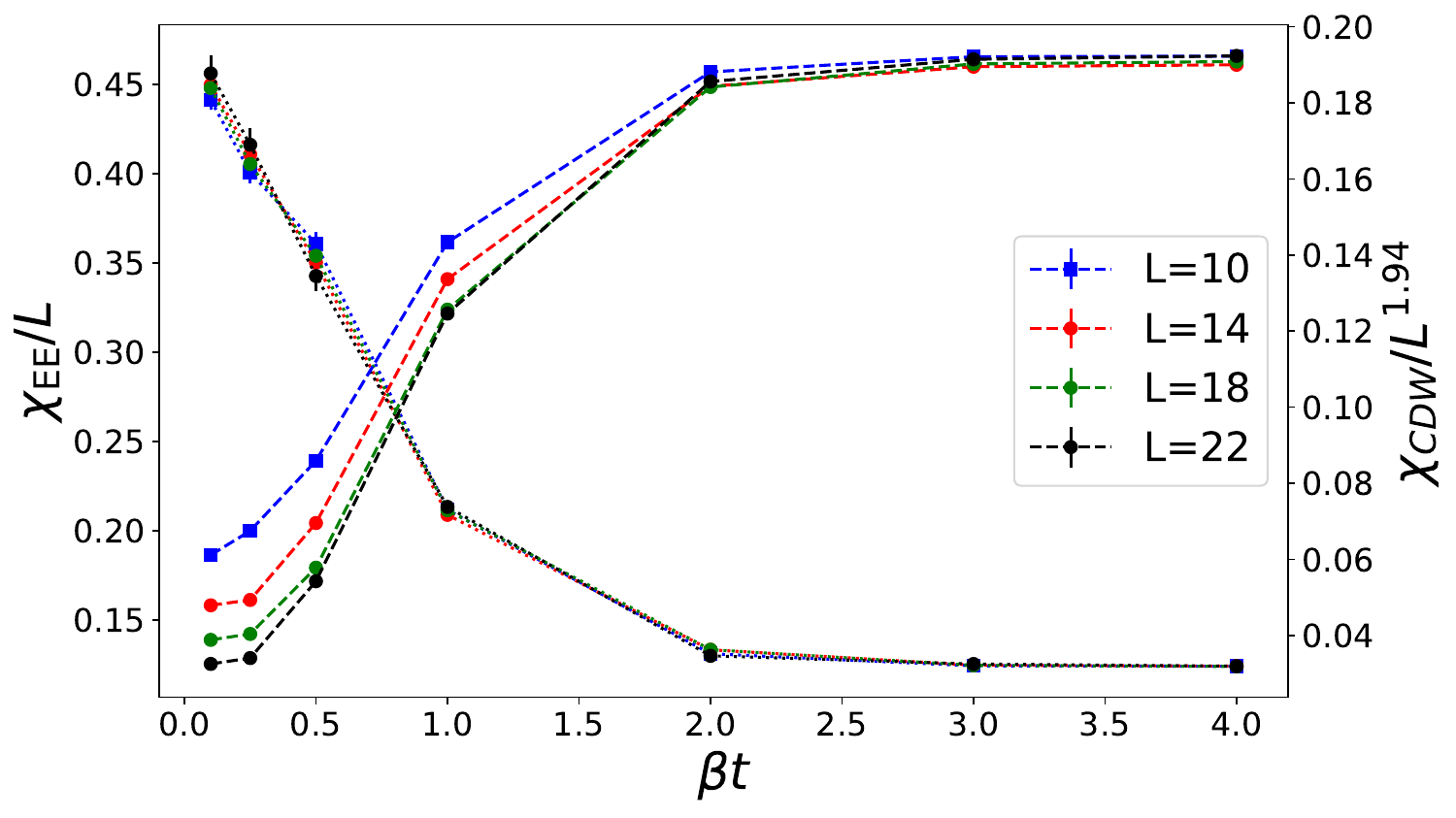}
    \caption{Finite temperature data for $U(1)$ theory in $1+1$-$d$. The dotted lines show the $\chi_{EE}$, which is
    the susceptibility corresponding to ${\cal E}$. This value rapidly converges to 0.125. On the other hand, the dashed
    lines trace the $\chi_{CDW}$ which display somewhat more finite size effects. The thermal behaviour of both observables 
    indicate that the transition from low to high temperature is associated with a smooth crossover, consistent
    with the literature.}
    \label{fig:transitionsup}
\end{figure}

 First, in Fig. \ref{fig:transitionsup} we show the behaviour of the normalized susceptibilities corresponding 
 to ${\cal E}$ and the CDW operator as a function of temperature for smaller lattices up to $L=22$, filtered to 
 include only the $G_x=0$ sector (an exponential effort for high temperatures). While they converge to their zero 
 temperature values relatively quickly as a function of the volume, we are also able to capture the finite 
 temperature crossover.

 In the main text we have shown the probability distribution of $\bar{\psi} \psi$ which shows a double 
peaked structure. We emphasize that each of the peaks correspond to one of the two Gauss' Law sectors 
which emerge at low temperatures, indicating broken chiral symmetry. However, a corresponding histogram for the 
${\cal E}$, the order parameter for the $CP$ symmetry (see \cref{fig:EEhist} (left)) shows a symmetric distribution. 
This indicates constraining dynamics between the fermions and gauge fields: with the fermions constrained
to occupy certain lattice sites at low temperatures, the gauge fields can fluctuate smoothly. This in turn means
that the susceptibility corresponding to the electric flux (normalized to unit volume) cancel the cross
terms (where $x \neq y$) and converges to the trivial value 0.125, contributed from the contact terms $(s_{x,x+1}^3)^2$. 
We can characterize the fermionic order by measuring the divergence of the susceptibility of the charge density 
wave as a function of spatial volume, $L$. As we show in \cref{fig:chiDiv}, the susceptibility diverges as 
$L^{1.94}$, for both $\beta t = 4$ and $\beta t = 8$. At high temperatures, on the other hand, the fermions
can hop much more, which due to Gauss' Law implies that the gauge fields are only allowed certain
values. This is manifested in \cref{fig:EEhist} (right) which shows the prominence of certain electric fluxes
while others are highly suppressed. Moreover, this also explains the behaviour of the electric flux
susceptibility in \cref{fig:transition} of the main text: at low temperatures there is cancellation 
between the smooth electric fluxes, while at high temperatures the finite number of ${\cal E}$ values 
can multiply coherently to give rise to a large value of $\chi_{\cal E}$.

\begin{figure}
    \centering
    \includegraphics[width=3in]{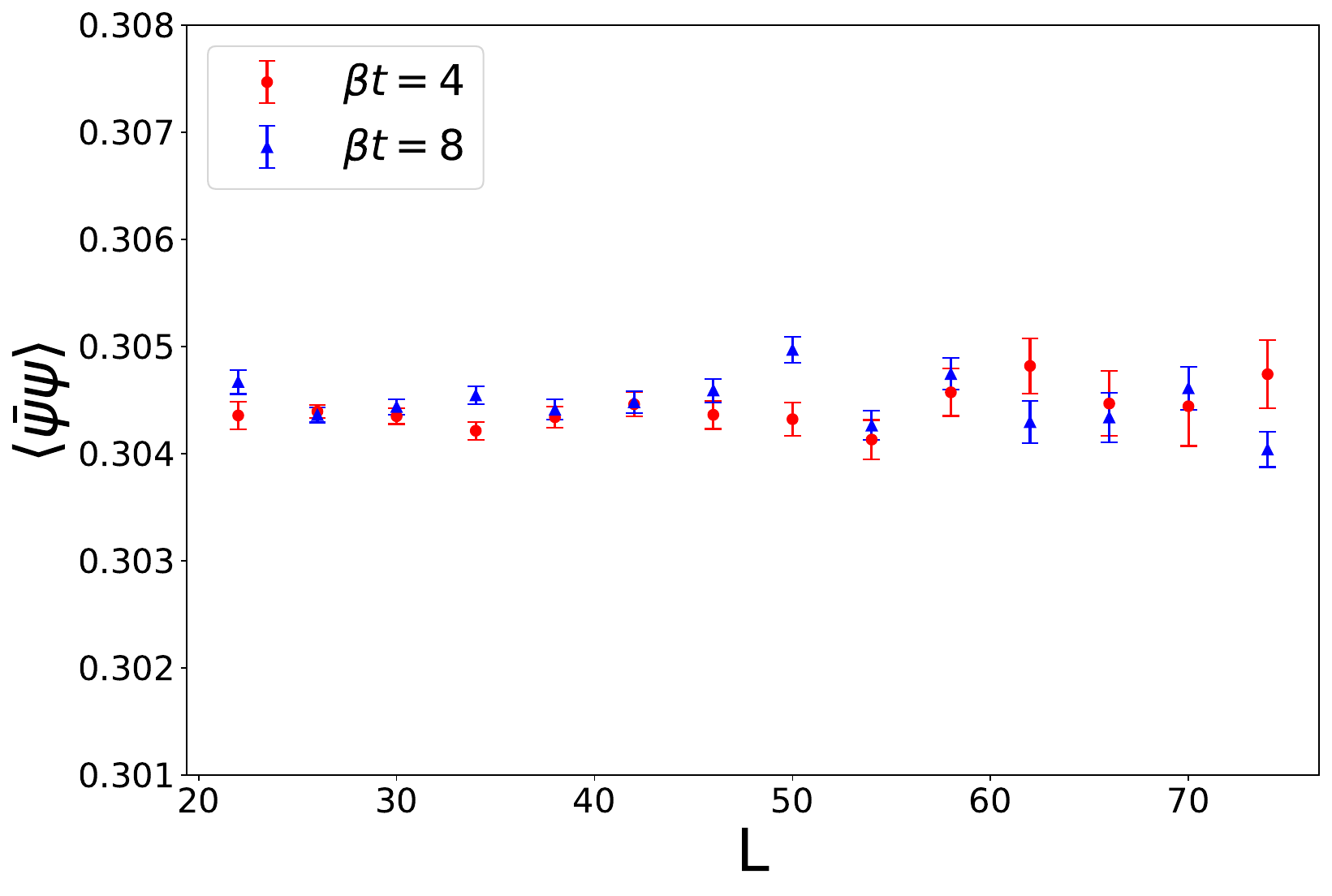}
    \includegraphics[width=3in]{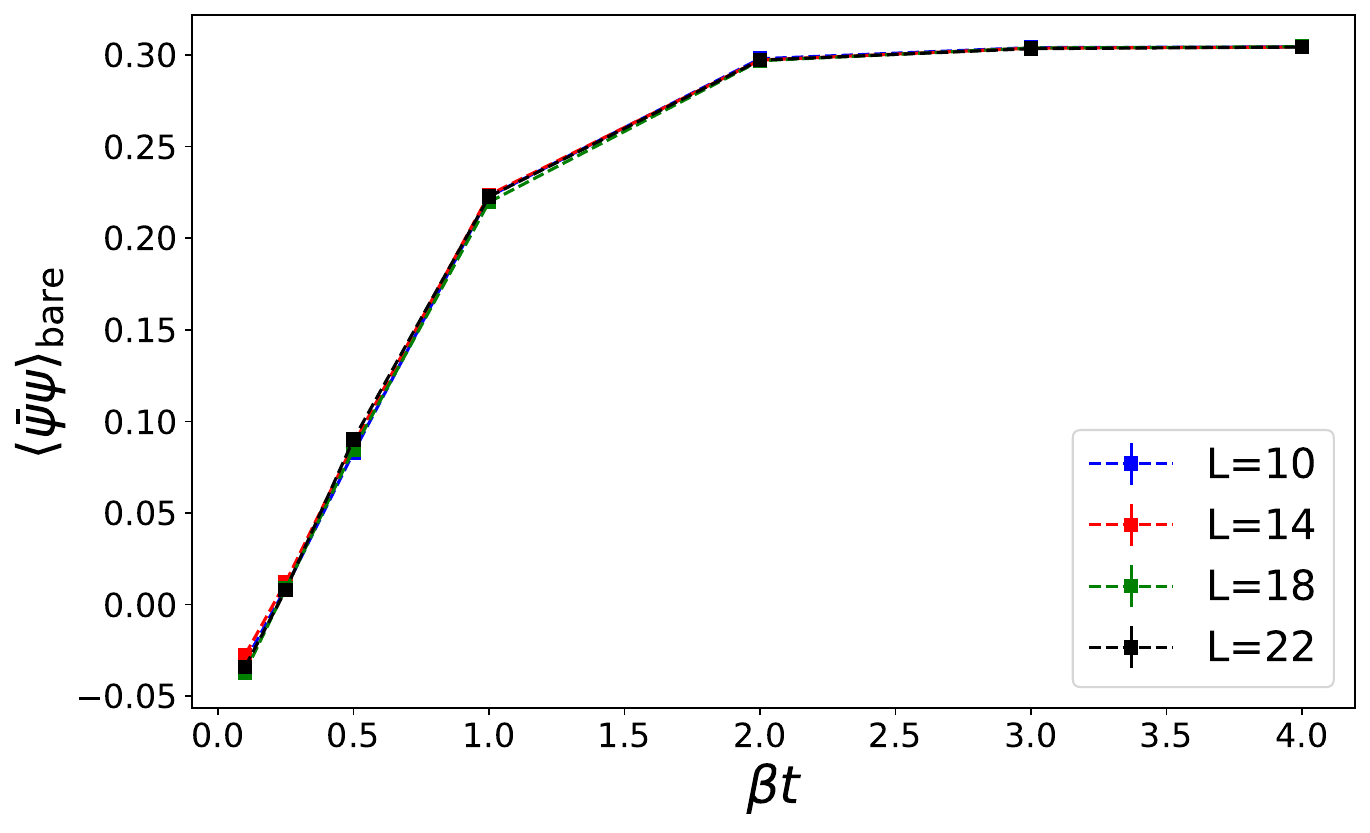}
    \caption{Bare chiral condensates for $U(1)$ theory in $1+1$-$d$ at low temperature and large lattices
    (top), while the temperature dependence of the condensate is bottom part of the figure.}
    \label{fig:chicond}
\end{figure}

 It is also interesting to note that the presence of the large CDW order also indicates that the model 
develops a large chiral condensate $\bar{\psi} \psi$. For the model with staggered fermions, it is known
that Gauss' Law explicitly breaks the $\mathbb{Z}_2$ chiral symmetry of the model, which consists of a single 
spatial lattice translation \cite{Banerjee:2012pg}. The chiral condensate is thus large at small temperatures
and as one increases the temperature, the condensate smoothly decreases and is expected to vanish at 
large temperatures. Since this is a crossover, there is no sharp behaviour in the condensate. Moreover,
due to this explicit breaking of the $\mathbb{Z}_2$ symmetry by the Gauss Law, the chiral condensate
undergoes additive renormalization in the Schwinger model. A similar phenomenology also happens in our
model, but since we have a four-fermi coupling instead of a mass term, it is non-trivial to 
match the two models quantitatively without doing an extensive program involving non-perturbative
renormalization. Nevertheless, on comparing the bare chiral condensate at low temperatures on large 
lattices in the thermodynamic limit, with the corresponding values quoted for the Schwinger model in 
Fig 15 of \cite{Byrnes:2002nv}, we estimate that our model is equivalent to the Kogut-Susskind Schwinger 
model on the lattice with a bare mass of $a m \sim 0.38 - 0.40$. In \cref{fig:chicond} (top), we 
show how the chiral condensate approaches a smooth thermodynamic limit. In the bottom panel
of \cref{fig:chicond}, we also show the measurement of the chiral condensate with increasing temperatures
until the condensate vanishes (and even becomes slightly negative at infinite temperatures), indicating
a restoration of chiral symmetry at high temperatures. While it is qualitatively the same scenario in the
Schwinger model \cite{Banuls:2016lkq}, we refrain from getting into a technical discussion about the 
technical aspects of the renormalized chiral condensates in this paper. 
Moreover, unlike the Schwinger model the $CP$ symmetry is not spontaneously broken
in the phase with $V=2t$. This leaves open the possibility of accessing the quantum phase transition
where $CP$ gets spontaneously broken by tuning other bare parameters in the Hamiltonian.

\end{document}